# To the problem of turbulence in quantitative easing transmission channels and transactions network channels at quantitative easing policy implementation by central banks

Dimitri O. Ledenyov and Viktor O. Ledenyov

*Abstract* – The central banks introduced a series of quantitative easing programs and decreased the long term interest rates to near zero with the aim to ease the credit conditions and provide the liquidity into the financial systems, responding to the 2007-2013 financial crisis in the USA, UK, Western Europe, and Japan. We review the U.S. Federal Reserve System, European Central bank, Bank of England and Bank of Japan monetary and financial policies with the particular focus on the quantitative easing policy implementation in the USA. Discussing some aspects of the quantitative easing policy implementation, we highlight the fact that the levels of capital change quickly in both the quantitative easing transmission channels and the transaction networks channels during the quantitative easing policy implementation, when the liquidity is added to the financial system. In agreement with the recent research findings in the econophysics, we propose that the nonlinear dynamic chaos can be generated by the turbulent capital flows in both the quantitative easing transmission channels and the transaction networks channels, when there are the laminar - turbulent capital flow transitions in the financial system. We demonstrate that the capital flows in both the quantitative easing transmission channels and the transaction networks channels in the financial system can be accurately characterized by the Reynolds numbers. We explain that the transition to the nonlinear dynamic chaos regime can be realized through the cascade of the Landau – Hopf bifurcations in the turbulent capital flows in both the quantitative easing transmission channels and the transaction networks channels in the financial system. Finally, we clarify that the general approach to the modeling of the US economy is based on both 1) the "large" empirically-motivated regression model, and 2) the Value at Risk (VAR) model with the incorporated Smets-Wouters model, which don't take to the account the origination of the nonlinear dynamic chaos regime during the capital flow in the quantitative easing transmission channels and in the transaction networks channels in the financial system. Therefore, we completed the computer modeling, using both the Nonlinear Dynamic Stochastic General Equilibrium Theory (NDSGET) and the Hydrodynamics Theory (HT), to accurately characterize the US economy in the conditions of the QE policy implementation by the US Federal Reserve. We found that the ability of the US financial system to adjust to the different levels of liquidity depends on the nonlinearities appearance in the QE transmission channels, and is limited by the laminar – turbulent capital flows transitions in the QE transmission channels and the transaction networks channels in the US financial system. The proposed computer model allows us to make the accurate forecasts of the US economy performance in the cases, when there are the different levels of liquidity in the US financial system.

# Introduction

"The practice of **monetary policy** has evolved a great deal since the early *1990s*. This evolution was significantly influenced by rapid developments in the theory of monetary policy. Inflation targeting has established itself as the dominant framework for monetary policy decisions," as explained in *Baltensperger, Hildebrand, Jordan (2007), Bernanke B S (1979 - 2013)*. For example, the monetary policy at the *SNB* includes the three-part monetary policy strategy in *Swiss National Bank Quarterly Bulletin (2013)*: "*1)* First, it regards prices as stable when the national consumer price index (*CPI*) rises by less than *2%* per annum. This allows it to take account of the fact that the *CPI* slightly overstates actual inflation. At the same time, it allows inflation to fluctuate somewhat with the economic cycle. *2)* Second, the *SNB* summarises its assessment of the situation and of the need for monetary policy action in a quarterly inflation forecast. This forecast, which is based on the assumption of a constant short-term interest rate, shows how the *SNB* expects the *CPI* to move over the next three years. *3)* Third, the *SNB* sets its operational goal in the form of a target range for the three-month *Swiss* franc *Libor*. In addition, a minimum exchange rate against the euro is currently in place." However, there is a strong need for the unconventional monetary policy introduction, because of the existing economic crisis in the Western Europe and North America (see the most recent data on the *GDP* in the *Switzerland, Euro area, U.S.A., U.K.* in Fig. 1 in *Swiss National Bank Financial Stability Report (2012)*, and the *U.S. Federal Reserve System, European Central bank, Bank of Japan and Bank of England* main interest rates in Fig. 2 in *Fawley, Neely (2013))*.

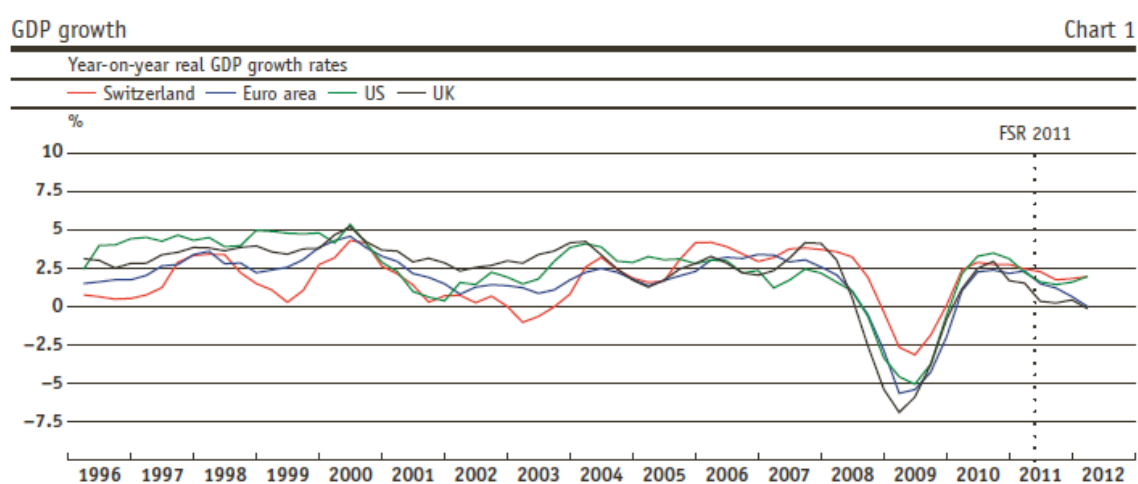

*Fig. 1. Gross Domestic Product (GDP) growth chart (after Swiss National Bank Financial Stability Report (2012)).*

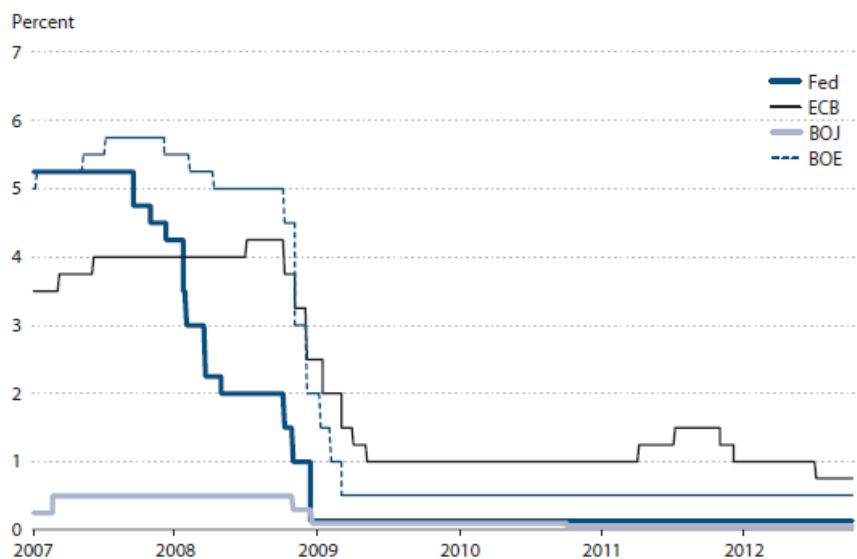

*Fig. 2.* U.S. Federal Reserve System, European Central bank, Bank of Japan and Bank of England main interest rates (after Fawley, Neely (2013)).

The true meaning of *quantitative easing* expression, introduced by *Werner R A* in *1995*, is to increase the net credit creation in the four possible ways in *Werner (2009)*:

1) by increasing bank credit,

2) by increasing trade credit,

3) by increasing central bank credit, and

4) by increasing credit created by the government.

The *Bank of Japan* used the *quantitative easing* expression to describe its monetary policy to increase the *Bank of Japan* reserves. The *Japanese* expression for the *quantitative easing* (量的金融緩和, *ryōteki kin'yū kanwa*) was frequently used by the *Bank of Japan* in *Japan* in *Bank of Japan (2001), Hiroshi Fujiki et al (2001), Shirakawa (2002)*. The concise definition of the *quantitative easing* policy is provided in *Wikipedia (2013)*: "**Quantitative Easing (QE) is an unconventional monetary policy used by central banks to stimulate the national economy when standard monetary policy has become ineffective** in *Bank of England (2011a)*. A central bank implements quantitative easing by buying financial assets from commercial banks and other private institutions, thus creating money and injecting a pre-determined quantity of money into the economy. This is distinguished from the more usual policy of buying or selling government bonds to change money supply, in order to keep market interest rates at a specified target value in *Bank of England (2011b)*. Expansionary monetary policy typically involves the central bank buying short-term government bonds in order to lower short-term market interest rates in *European Central Bank (2008)*. However, when short-term interest rates are either at, or close to, zero, normal monetary policy can no longer lower interest rates. *Quantitative easing* may then be used by the monetary authorities to further stimulate the economy by purchasing assets of longer maturity than only short-term government bonds, and thereby lowering longer-term interest rates further out on the yield curve in *Bernanke (2009)*. *Quantitative easing* raises the prices of the financial assets bought, which lowers their yield in *Larry (2009)*. *Quantitative easing* can be used to help ensure that inflation does not fall below target in *Bank of England (2011b)*. Risks include the policy being more effective than intended in acting against deflation – leading to higher inflation in *Bowlby (2009)* or of not being effective enough if banks do not lend out the additional reserves in *Isidore (2010)*. According to the *IMF* and various other economists, *quantitative easing* undertaken since the global financial crisis has mitigated the adverse effects of the crisis in *Klyuev, de Imus, Srinivasan (2009)*."

Let us sum up the above information by saying that, the *quantitative easing* policy introduction and implementation means that the central bank releases the sufficient capital to the commercial and investment banks in order to add the liquidity to the national banking system with the goal to improve the existing situation in the national economy.

Let us briefly review the present developments toward the *QE* policy implementation with the help of the *large-scale asset purchases* (*LSAPs*) in the *U.S.A.* in *Christensen, Rudebusch (2012)*, *D'Amico, King (2011), Gagnon, Raskin, Remache, Sack (2011)*. The *U.S. Federal Reserve System* held between *US$700* billion and *US$800* billion of *Treasury* notes on its balance sheet before the recession. Presently, the *U.S. Federal Reserve System* holds around *US$2.5-3.0* trillion of bank debt, mortgage-backed securities, and Treasury notes in *Wikipedia (2013)*. The *U.S. Federal Reserve System* conducted the following *QE* programs in *Wikipedia (2013)*:

1. *QE1*: the purchase of *US$600* billion in the mortgage-backed securities in 2008 - 2010;

2. *QE2*: the purchase of *US$600* billion in the treasury securities in 2010 - 2011;

3. *QE3*: the purchase of *US$40* billion of mortgage-backed securities (*MBS*) per month since *September, 2012* until *2015*.

*Christensen, Rudebusch (2012), Fawley, Neely (2013)* summarized the information on some of the *U.S. Federal Reserve System QE* programs announcements in Tabs. 1, 2. The *U.S. Federal Reserve bank assets* are shown in Fig. 3 in *Fawley, Neely (2013)*.

| No. | Date | Event | Description |
|-----|------|-------|-------------|
| I | Nov. 25, 2008 | Initial LSAP announcement | Fed announces purchases of $100 billion in GSE debt and up to $500 billion in MBS. |
| II | Dec. 1, 2008 | Bernanke speech | Chairman Bernanke indicates that the Fed could purchase long-term Treasury securities. |
| III | Dec. 16, 2008 | FOMC statement | The first FOMC statement that mentions possible purchases of long-term Treasuries. |
| IV | Jan. 28, 2009 | FOMC statement | FOMC states that it is ready to expand agency debt and MBS purchases and to purchase long-term Treasuries. |
| V | Mar. 18, 2009 | FOMC statement | Fed will purchase an additional $750 billion in agency MBS and $100 billion in agency debt. Also, it will purchase $300 billion in long-term Treasury securities. |
| VI | Aug. 12, 2009 | FOMC statement | Fed is set to slow the pace of the LSAP. The final purchases of Treasury securities will be in the end of October instead of mid-September. |
| VII | Sep. 23, 2009 | FOMC statement | Fed's purchases of agency debt and MBS will end in the first quarter of 2010, while its Treasury purchases will end as planned in October. |
| VIII | Nov. 4, 2009 | FOMC statement | Amount of agency debt capped at $175 billion instead of the $200 billion previously announced. |

*Tab. 1.* U.S. Federal Reserve System Key QE Announcements

(after Christensen, Rudebusch (2012)).

**Important Announcements by the Federal Reserve**

| Date | Program | Event | Brief description | Interest rate news |
|------|---------|-------|-------------------|---------------------|
| 11/25/2008 | QE1 | FOMC statement | LSAPs announced: Fed will purchase $100 billion in GSE debt and $500 billion in MBS. | |
| 12/1/2008 | QE1 | Bernanke speech | First suggestion of extending QE to Treasuries. | |
| 12/16/2008 | QE1 | FOMC statement | First suggestion of extending QE to Treasuries by FOMC. | The Fed cuts the federal funds rate from 1% to 0.00-0.25%; expects low rates "for some time." |
| 1/28/2009 | QE1 | FOMC statement | Fed stands ready to expand QE and buy Treasuries. | |
| 3/18/2009 | QE1 | FOMC statement | LSAPs expanded: Fed will purchase $300 billion in long-term Treasuries and an additional $750 and $100 billion in MBS and GSE debt, respectively. | Fed expects low rates for "an extended period." |
| 8/12/2009 | QE1 | FOMC statement | LSAPs slowed: All purchases will finish by the end of October, not mid-September. | |
| 9/23/2009 | QE1 | FOMC statement | LSAPs slowed: Agency debt and MBS purchases will finish at the end of 2010:Q1. | |
| 11/4/2009 | QE1 | FOMC statement | LSAPs downsized: Agency debt purchases will finish at $175 billion. | |
| 8/10/2010 | QE1 | FOMC statement | Balance sheet maintained: The Fed will reinvest principal payments from LSAPs in Treasuries. | |
| 8/27/2010 | QE2 | Bernanke speech | Bernanke suggests role for additional QE "should further action prove necessary." | |
| 9/21/2010 | QE2 | FOMC statement | FOMC emphasizes low inflation, which "is likely to remain subdued for some time before rising to levels the Committee considers consistent with its mandate." | |
| 10/12/2010 | QE2 | FOMC minutes released | FOMC members' "sense" is that "[additional] accommodation may be appropriate before long." | |
| 10/15/2010 | QE2 | Bernanke speech | Bernanke reiterates that Fed stands ready to further ease policy. | |
| 11/3/2010 | QE2 | FOMC statement | QE2 announced: Fed will purchase $600 billion in Treasuries. | |
| 6/22/2011 | QE2 | FOMC statement | QE2 finishes: Treasury purchases will wrap up at the end of month, as scheduled; principal payments will continue to be reinvested. | |
| 9/21/2011 | Maturity Extension Program | FOMC statement | Maturity Extension Program ("Operation Twist") announced: The Fed will purchase $400 billion of Treasuries with remaining maturities of 6 to 30 years and sell an equal amount with remaining maturities of 3 years or less; MBS and agency debt principal payments will no longer be reinvested in Treasuries, but instead in MBS. | |
| 6/20/2012 | Maturity Extension Program | FOMC statement | Maturity Extension Program extended: The Fed will continue to purchase long-term securities and sell short-term securities through the end of 2012. Purchases/sales will continue at the current pace, about $45 billion/month. | |
| 8/22/2012 | QE3 | FOMC minutes released | FOMC members "judged that additional monetary accommodation would likely be warranted fairly soon…" | |
| 9/13/2012 | QE3 | FOMC statement | QE3 announced: The Fed will purchase $40 billion of MBS per month as long as "the outlook for the labor market does not improve substantially…in the context of price stability." | Fed expects low rates "at least through mid-2015." |
| 12/12/2012 | QE3 | FOMC statement | QE3 expanded: The Fed will continue to purchase $45 billion of long-term Treasuries per month but will no longer sterilize purchases through the sale of short-term Treasuries. | The Fed expects low rates to be appropriate while unemployment is above 6.5 percent and inflation is forecasted below 2.5 percent. |

*Tab. 2.* U.S. Federal Reserve System Important QE Announcements

(after Fawley, Neely (2013)).

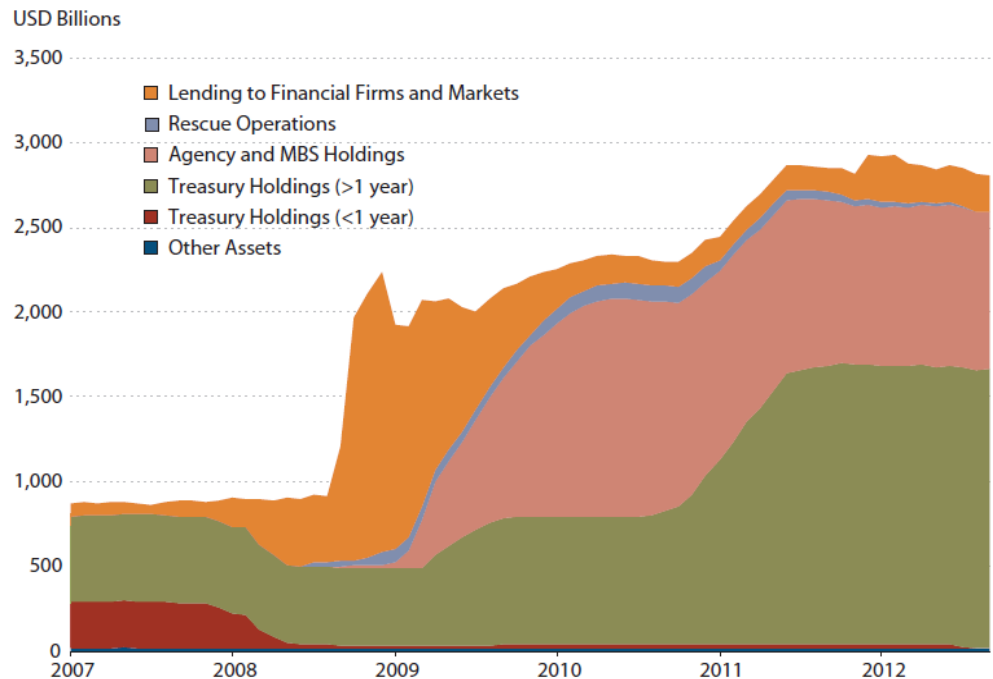

***Fig. 3.*** *U.S. Federal Reserve bank assets (after Fawley, Neely (2013)).*

Let us briefly summarize the present developments as far as the *QE* policy implementation is concerned in the *U.K* in *Benford, Berry, Nikolov, Young (2009)*, *Fisher (2010a, 2010b)*, *Joyce, Tong, Woods (2011), Bridges, Rossiter, Thomas (2011), Joyce, Lasaosa, Stevens, Tong (2011)*. The main completed *QE* programs by the *Bank of England* are listed below in *Wikipedia (2013)*:

1. The *Bank of England* had purchased around *£165* billion of assets in *2008 - 2009*.
2. The *Bank of England* had increased the total asset purchases up to *£175* billion in *2009 – 2010*.
3. The *Bank of England* had increased the total asset purchases up to *£250* billion as of *2011*.
4. The *Bank of England* had increased the total asset purchases up to *£375* billion in *2011 - 2012*.

*Christensen, Rudebusch (2012), Fawley, Neely (2013), Joyce, Tong, Woods (2011)* summarized some of the *Bank of England QE* programs announcements in Tabs. 3, 4 , 5. The *Bank of England* assets are shown in Fin. 4 in *Fawley, Neely (2013)*.

| No. | Date | Event | Description |
|---|---|---|---|
| I | Feb. 11, 2009 | February Inflation Report | Press conference and Inflation Report indicated that asset purchases were likely. |
| II | Mar. 5, 2009 | MPC statement | The MPC announced that it would purchase £75 billion of assets over three months. Gilt purchases would be restricted to the 5-25 year maturity range. |
| III | May 7, 2009 | MPC statement | The MPC announced that the amount of asset purchases would be extended by a further £50 billion to a total of £125 billion. |
| IV | Aug. 6, 2009 | MPC statement | The MPC announced that the amount of asset purchases would be extended to £175 billion and that the buying range would be extended to include gilts with residual maturity greater than three years. |
| V | Nov. 5, 2009 | MPC statement | The MPC announced that the asset purchases would be extended to £200 billion. |
| VI | Feb. 4, 2010 | MPC statement | The MPC announced that the amount of asset purchases would be maintained at £200 billion. |
| VII | Oct 6, 2011 | MPC statement | The MPC announced that the asset purchases would be extended to £275 billion. |

***Tab. 3.*** *Bank of England key QE announcements (after Christensen, Rudebusch (2012)).*

**Important Announcements by the Bank of England**

| Date | Program | Event | Brief description | Interest rate news |
|---|---|---|---|---|
| 1/19/2009 | APF | HM Treasury statement | APF established: The BOE will purchase up to £50 billion of "high quality private sector assets" financed by Treasury issuance. | |
| 2/11/2009 | APF | BOE Inflation Report released | The BOE views a slight downside risk to meeting the inflation target, reiterates APF as a potential policy instrument. | |
| 3/5/2009 | APF | MPC statement | QE announced: The BOE will purchase up to £75 billion in assets, now financed by reserve issuance; medium- and long-term gilts will comprise the "majority" of new purchases. | The BOE cuts policy rate from 1% to 0.5%; the ECB cuts policy rate from 2% to 1.5%. |
| 5/7/2009 | APF | MPC statement | QE expanded: The BOE will purchase up to £125 billion in assets. | |
| 8/6/2009 | APF | MPC statement | QE expanded: The BOE will purchase up to £175 billion in assets; to accommodate the increased size, the BOE will expand purchases into gilts with remaining maturity of 3 years or more. | |
| 11/5/2009 | APF | MPC statement | QE expanded: The BOE will purchase up to £200 billion in assets. | |
| 2/4/2010 | APF | MPC statement | QE maintained: The BOE maintains the stock of asset purchases financed by the issuance of reserves at £200 billion; new purchases of private assets will be financed by Treasury issuance. | |
| 10/6/2011 | APF | MPC statement | QE expanded: The BOE will purchase up to £275 billion in assets financed by reserve issuance; the ceiling on private assets held remains £50 billion. | |
| 11/29/2011 | APF | HM Treasury decision | Maximum private asset purchases reduced: HM Treasury lowers the ceiling on APF private asset holdings from £50 billion to £10 billion. | |
| 2/9/2012 | APF | MPC statement | QE expanded: The BOE will purchase up to £325 billion in assets. | |
| 7/5/2012 | APF | MPC statement | QE expanded: The BOE will purchase up to £375 billion in assets. | |

NOTE: MPC, Monetary Policy Committee.

***Tab. 4.*** *Bank of England important QE announcements (after Fawley, Neely (2013)).*

| Date | Event |
|---|---|
| **2009** | |
| 19 January | The Chancellor of the Exchequer announces that the Bank of England will set up an asset purchase programme. |
| 30 January | Asset Purchase Facility Fund established. Exchange of letters between the Chancellor of the Exchequer and the Governor on 29 January 2009. |
| 5 February | Bank Rate reduced from 1.5% to 1%. |
| 11 February | February *Inflation Report* and the associated press conference give strong indication that QE asset purchases are likely. |
| 13 February | First purchases of commercial paper begin. |
| 5 March | Bank Rate reduced from 1% to 0.5%. The MPC announces it will purchase £75 billion of assets over three months funded by central bank money. Conventional bonds likely to constitute the majority of purchases, restricted to bonds with residual maturity between 5 and 25 years. |
| 11 March | First purchases of gilts begin. |
| 25 March | First purchases of corporate bonds begin. |
| 7 May | The MPC announces that the amount of QE asset purchases will be extended by a further £50 billion to £125 billion. |
| 3 August | Secured commercial paper facility launched. |
| 6 August | The MPC announces that QE asset purchases will be extended to £175 billion and that the buying range will be extended to gilts with a residual maturity greater than three years. The Bank announces a gilt lending programme, which allows counterparties to borrow gilts from the APF's portfolio via the DMO in return for a fee and alternative gilts as collateral. |
| 5 November | The MPC announces that QE asset purchases will be extended to £200 billion. |
| 22 December | The Bank announces that it will act as a seller, as well as a buyer, of corporate bonds in the secondary market. |
| **2010** | |
| 8 January | First sales of corporate bonds. |
| 4 February | The MPC announces that QE asset purchases will be maintained at £200 billion. The Chancellor authorises the Bank to continue to transact in private sector assets, with further purchases financed by issuance of Treasury bills. The MPC's press statement says that the Committee will continue to monitor the appropriate scale of the asset purchase programme and that further purchases will be made should the outlook warrant them. |

Shading denotes announcements used in the event study analysis by Joyce *et al* (2011) referred to in the following section.

***Tab. 5.*** *Bank of England key QE announcements (after Joyce, Tong, Woods (2011)).*

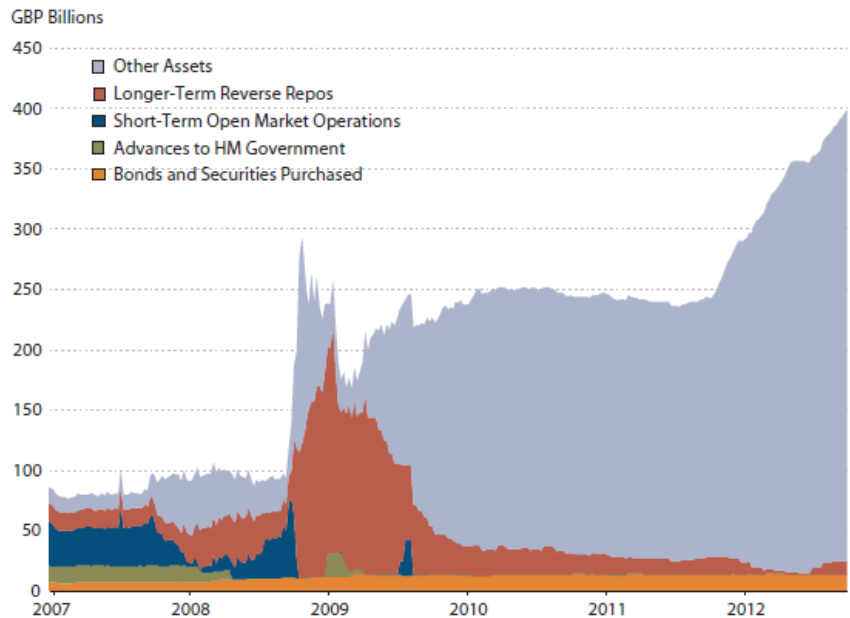

***Fig. 4.*** *Bank of England assets (after Fawley, Neely (2013)).*

Let us take a close look on the *QE* policy implementation in *Japan*. The completed *QE* programs by the *Bank of Japan (BOJ)* are listed below in *Bank of Japan (2001), Hiroshi Fujiki et al (2001), Shirakawa (2002), Wikipedia (2013)*:

1. The *Bank of Japan (BOJ)* increased the commercial bank current account balance from ¥5 trillion Yen to ¥*35* trillion Yen (approximately *US$300* billion) over a *4*-year period starting in March, *2001-2004*.

2. The *Bank of Japan (BOJ)* announced that it would purchase of ¥*5* trillion Yen (*US$60* billion) in assets in early *October, 2010*.

3. The *Bank of Japan (BOJ)* announced a unilateral move to increase the amount from ¥*40* trillion Yen (*US$504* billion) to a total of ¥*50* trillion Yen (*US$630* billion) on *August 4, 2011*.

4. The *Bank of Japan (BOJ)* expanded its asset purchase program by ¥*5* trillion Yen (*US$66* billion) to a total of ¥*55* trillion *Yen* in *October, 2011*.

*Fawley, Neely (2013)* analyzed the *Bank of Japan* important *QE announcements* in *Tab. 6;* and provided information on the *Bank of Japan* assets in Fig. 5.

**Important Announcements by the Bank of Japan**

| Date | Program | Event | Brief description | Interest rate news |
|---|---|---|---|---|
| 12/2/2008 | SFSOs | Unscheduled monetary policy meeting | The BOJ will operate a facility through the end of April to lend an unlimited amount to banks at the uncollateralized overnight call rate and collateralized by corporate debt. | |
| 12/19/2008 | Outright JGB/CFI purchases | Statement on monetary policy | Outright purchases expanded: The BOJ increases monthly JGB purchases (last increased October 2002) from ¥1.2 trillion to ¥1.4 trillion; they will also look into purchasing commercial paper. | The BOJ lowers the target for the uncollateralized overnight call rate from 0.3% to 0.1%. |
| 1/22/2009 | Outright CFI purchases | Statement on monetary policy | Outright purchases announced: The BOJ will purchase up to ¥3 trillion in commercial paper and ABCP and is investigating outright purchases of corporate bonds. | |
| 2/19/2009 | Outright CFI purchases | Statement on monetary policy | Outright purchases expanded: The BOJ will extend commercial paper purchases and the SFSOs through the end of September (previously end of March) and will purchase up to ¥1 trillion in corporate bonds. | |
| 3/18/2009 | Outright JGB purchases | Statement on monetary policy | Outright purchases expanded: The BOJ increases monthly JGB purchases from ¥1.4 trillion to ¥1.8 trillion. | |
| 7/15/2009 | Outright CFI purchases/SFSOs | Statement on monetary policy | Programs extended: The BOJ extends the SFSOs and outright purchases of corporate paper and bonds through the end of the year. | |
| 10/30/2009 | Outright CFI purchases/SFSOs | Statement on monetary policy | Status of programs: Outright purchases of corporate finance instruments will expire at the end of 2009 as expected, but the SFSOs will be extended through 2010:Q1; ample liquidity provision past 2010:Q1 will occur through funds-supplying operations against pooled collateral, which will accept a larger range of collateral. | |
| 12/1/2009 | FROs | Statement on monetary policy | Facility announcement: The BOJ will offer ¥10 trillion in 3-month loans against the full menu of eligible collateral at the uncollateralized overnight call rate. | |
| 3/17/2010 | FROs | Statement on monetary policy | Facility expansion: The BOJ expands the size of the FROs to ¥20 trillion. | |
| 5/21/2010 | GSFF | Statement on monetary policy | GSFF announcement: The BOJ will offer ¥3 trillion in 1-year loans to private financial institutions with project proposals for "strengthening the foundations for economic growth." | |
| 8/30/2010 | FROs | Unscheduled monetary policy meeting | Facility expansion: The BOJ adds ¥10 trillion in 6-month loans to the FROs. | |
| 10/5/2010 | CME | Statement on monetary policy | APP established: The BOJ will purchase ¥5 trillion in assets (¥3.5 trillion in JGBs and Treasury discount bills, ¥1 trillion in commercial paper and corporate bonds, and ¥0.5 trillion in ETFs and J-REITs). | The BOJ sets the target for the uncollateralized overnight call rate at around 0 to 0.1%. |

| Date | Program | Event | Brief description | Interest rate news |
|---|---|---|---|---|
| 3/14/2011 | CME | Statement on monetary policy | APP expanded: The BOJ will purchase an additional ¥5 trillion in assets (¥0.5 trillion in JGBs, ¥1 trillion in Treasury discount bills, ¥1.5 trillion in commercial paper, ¥1.5 trillion in corporate bonds, ¥0.45 trillion in ETFs, and ¥0.05 trillion in J-REITs). | |
| 6/14/2011 | GSFF | Statement on monetary policy | GSFF expanded: The BOJ makes available another ¥0.5 trillion in loans to private financial institutions for the purpose of investing in equity and extending asset-based loans. | |
| 8/4/2011 | CME | Statement on monetary policy | APP/FROs expanded: The BOJ will purchase an additional ¥5 trillion in assets (¥2 trillion in JGBs, ¥1.5 trillion in Treasury discount bills, ¥0.1 trillion in commercial paper, ¥0.9 trillion in corporate bonds, ¥0.5 trillion in ETFs, and ¥0.01 trillion in J-REITs); 6-month collateralized loans through the FROs are expanded by ¥5 trillion. | |
| 10/27/2011 | CME | Statement on monetary policy | APP expanded: The BOJ will purchase an additional ¥5 trillion in JGBs. | |
| 2/14/2012 | CME | Statement on monetary policy | APP expanded: The BOJ will purchase an additional ¥10 trillion in JGBs. | |
| 3/13/2012 | GSFF | Statement on monetary policy | GSFF expanded: The BOJ makes available another ¥2 trillion in loans to private financial institutions, including ¥1 trillion in U.S.-dollar-denominated loans and ¥0.5 trillion in smaller-sized (¥1 million-¥10 million) loans. | |
| 4/27/2012 | CME | Statement on monetary policy | APP expanded/FROs reduced: The BOJ will purchase an additional ¥10 trillion in JGBs, ¥0.2 trillion in ETFs, and ¥0.01 trillion in J-REITs. The BoJ also reduces the availability of 6-month FRO loans by ¥5 trillion. | |
| 7/12/2012 | CME | Statement on monetary policy | APP expanded/FROs reduced: The BOJ will purchase an additional ¥5 trillion in Treasury discount bills and reduces the availability of FRO loans by ¥5 trillion. | |
| 9/19/2012 | CME | Statement on monetary policy | APP expanded: The BOJ will purchase an additional ¥5 trillion in JGBs and ¥5 trillion in Treasury discount bills. | |
| 10/30/2012 | CME/SBLF | Statement on monetary policy | APP expanded/SBLF announced: The BOJ will purchase an additional ¥5 trillion in JGBs, ¥5 trillion in Treasury discount bills, ¥0.1 trillion in commercial paper, ¥0.3 trillion in corporate bonds, ¥0.5 trillion in ETFs, and ¥0.01 trillion in J-REITs. Through the SBLF it will fund up to 100 percent of depository institutions' net increase in lending to the nonfinancial sector. | |
| 12/20/2012 | CME | Statement on monetary policy | APP expanded: The BOJ will purchase an additional ¥5 trillion JGBs and ¥5 trillion in Treasury discount bills. | |

NOTE: CFI, corporate finance instruments (corporate bonds plus commercial paper); CME, comprehensive monetary easing.

*Tab. 6.* Bank of Japan important QE announcements (after Fawley, Neely (2013)).

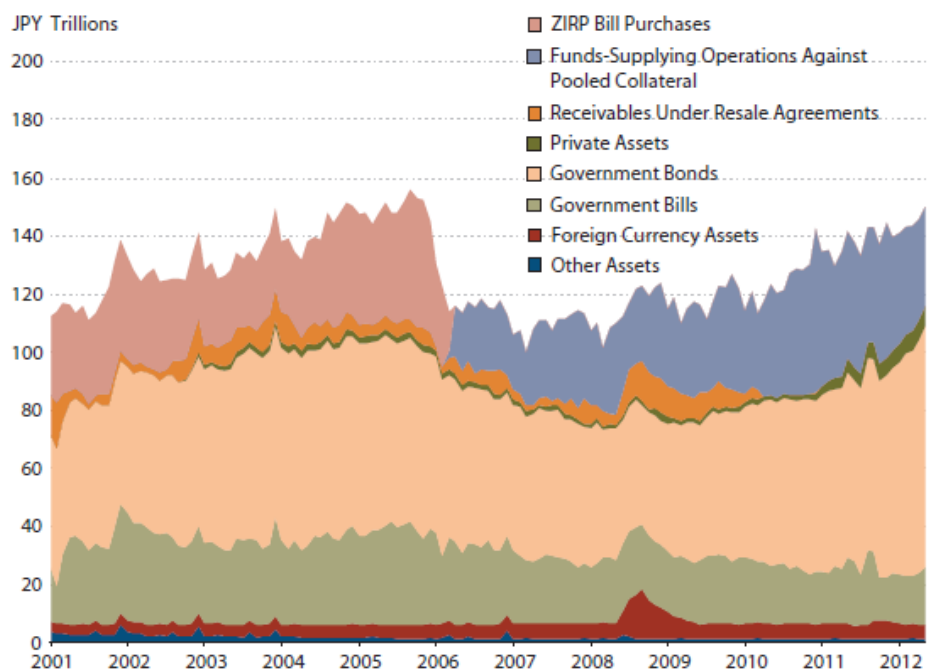

*Fig. 5.* Bank of Japan assets (after Fawley, Neely (2013)).

*Berkmen (2012)* analyzed the *Bank of Japan* current account balance and balance sheet as shown in Figs. 6 and 7.

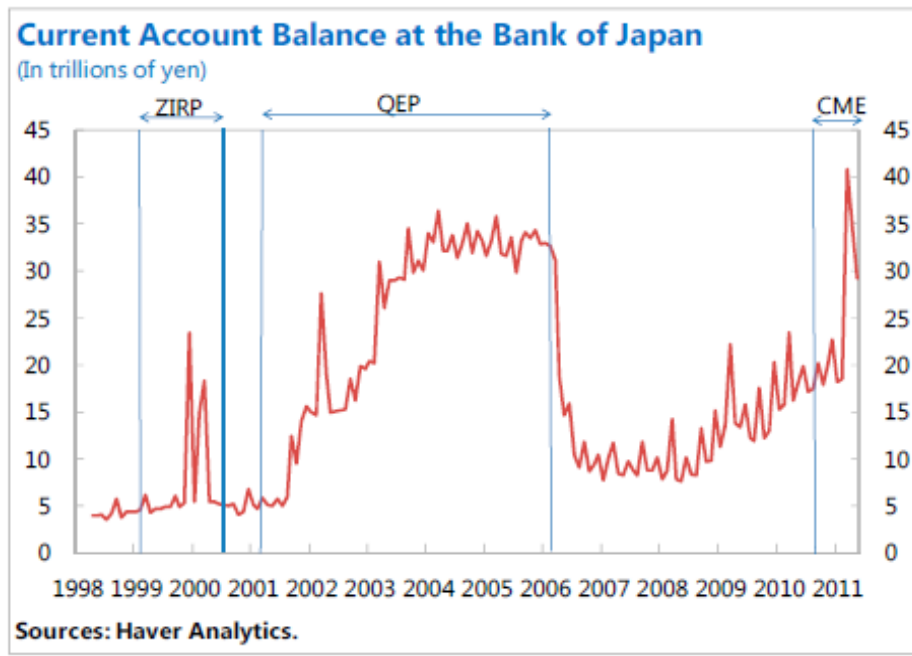

***Fig. 6.*** *Bank of Japan current account balance (after Berkmen (2012)).*

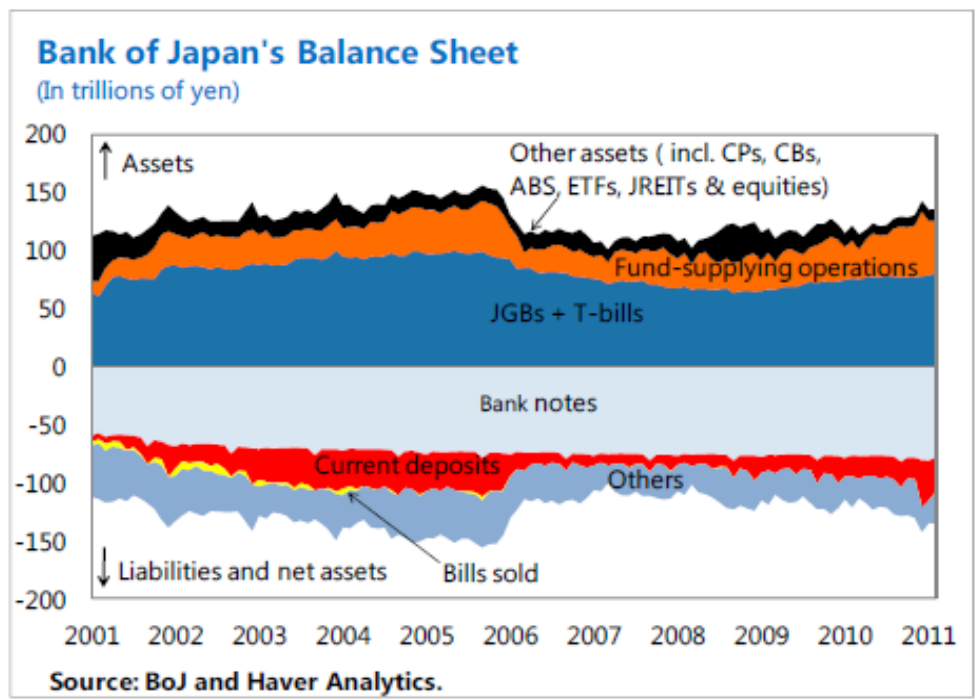

***Fig. 7.*** *Bank of Japan balance sheet (after Berkmen (2012)).*

*Wieland (2010)* researched some issues in the *QE* policy introduction and implementation in *Japan*, presenting the characteristic dependences in the Figs. 8, 9 and 10.

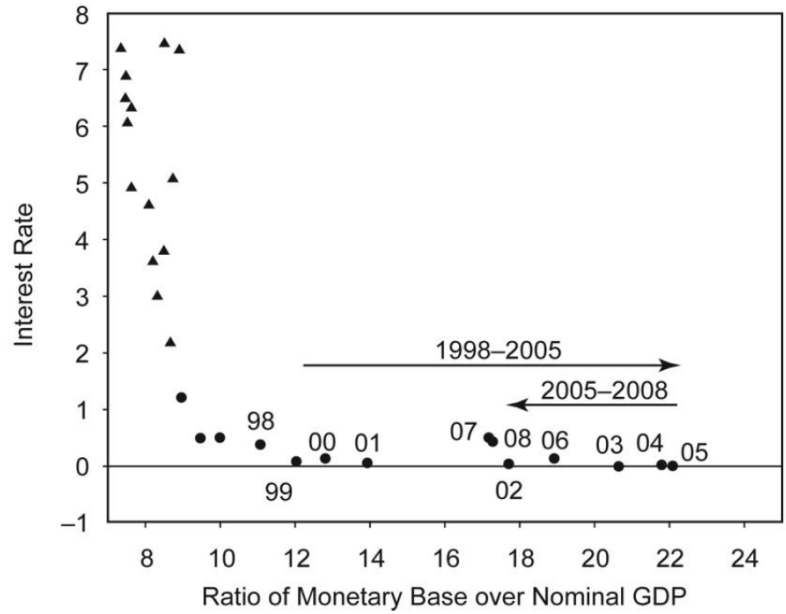

***Fig. 8.*** *The Marshallian k and the money market rate in Japan, 1981–2008, annual observations (after Wieland (2010)).*

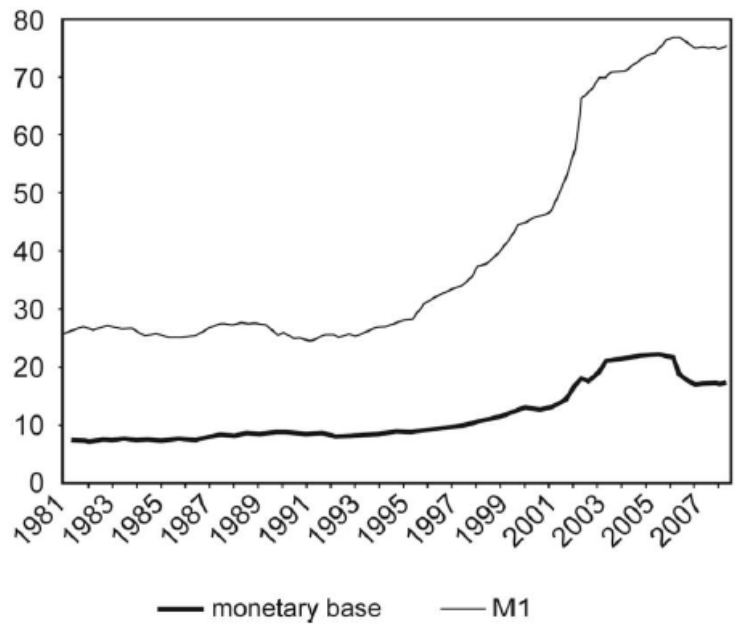

***Fig. 9.*** *The base money and M1 relative to nominal income in Japan, 1981–2008, quarterly observations (after Wieland (2010)).*

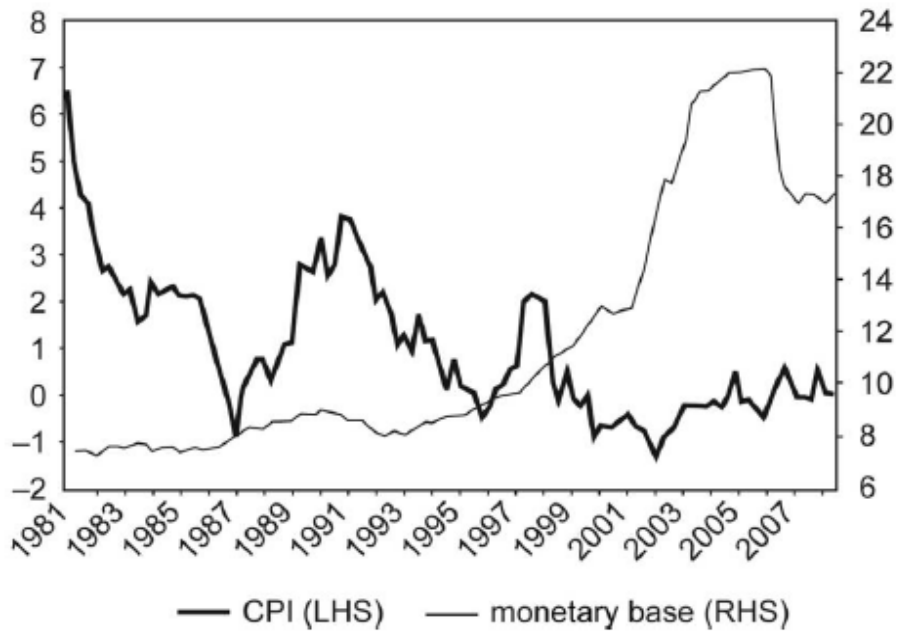

*Fig. 10. Base money and CPI inflation in Japan, 1981–2008, quarterly observations (after Wieland (2010)).*

*Assenmacher-Wesche, Gerlach, Sekine (2007)* also refer to an important remark, made by *Governor Mieno* in testimony to the *Diet* in *May 1987*: "We have paid attention to the developments of the money supply, as history not only in Japan but also in other countries suggests that in the long run a higher money growth rate tends to result in a higher inflation rate. Nevertheless, the relationship between money and inflation is much more complicated than a simple causal relationship implying that a certain rate of money growth will always result in a specified rise of inflation. For this reason, it is very difficult to determine the appropriate level of monetary aggregates and we need to rely on overall judgment based on all available information at the time."

*Assenmacher-Wesche, Gerlach, Sekine (2007)* came to the following important conclusion: "the band spectral regressions indicate that money growth is correlated – and output growth is inversely correlated – with inflation in the low frequency band, in particular when that is defined as frequencies of four years or more. Furthermore, that correlation reflects unidirectional *Granger* causality from money growth to inflation, implying that money growth does contain information about future inflation that is not already embedded in inflation. In the high frequency band the quantity-theoretic variables appear to be of little significance for inflation."

In Tab. 7, *the Bank of Japan* monetary base target and balance sheet projection are shown in Bank of Japan (2013).

|  | End-2012 (actual) | End-2013 (projected) | End-2014 (projected) |
|---|---|---|---|
| Monetary base | 138 | 200 | 270 |

Breakdown of the Bank's Balance Sheet

|  | End-2012 (actual) | End-2013 (projected) | End-2014 (projected) |
|---|---|---|---|
| JGBs | 89 | 140 | 190 |
| CP | 2.1 | 2.2 | 2.2 |
| Corporate bonds | 2.9 | 3.2 | 3.2 |
| Exchange-traded funds (ETFs) | 1.5 | 2.5 | 3.5 |
| Japan real estate investment trusts (J-REITs) | 0.11 | 0.14 | 0.17 |
| Loan Support Program | 3.3 | 13 | 18 |
| Total assets (including others) | 158 | 220 | 290 |
| Banknotes | 87 | 88 | 90 |
| Current deposits | 47 | 107 | 175 |
| Total liabilities and net assets (including others) | 158 | 220 | 290 |

(trillion yen)

*Tab. 7. Bank of Japan monetary base target and balance sheet projection (after Bank of Japan (2013)).*

*Fawley, Neely (2013)* researched the *QE* policy implementation by the *European Central Bank*, presenting the research data in Tab. 8; and showing the *European Central Bank assets* in Fig. 11.

**Important Announcements by the European Central Bank**

| Date | Program | Event | Brief description | Interest rate news |
|---|---|---|---|---|
| 3/28/2008 | LTRO | Governing Council press release | LTRO expanded: 6-month LTROs are announced. | |
| 10/15/2008 | FRFA | Governing Council press release | Refinancing operations expanded: All refinancing operations will be conducted with fixed-rate tenders and full allotment; the list of assets eligible as collateral in credit operations with the Bank is expanded to included lower-rated (with the exception of asset-backed securities) and non-euro-denominated assets. | |
| 5/7/2009 | CBPP/LTRO | Governing Council press release | CBPP announced/LTRO expanded: The ECB will purchase €60 billion in euro-denominated covered bonds; 12-month LTROs are announced. | ECB lowers the main refinancing rate by 0.25% to 1% and the rate on the marginal lending facility by 0.50% to 1.75%. |
| 5/10/2010 | SMP | Governing Council press release | SMP announced: The ECB will conduct interventions in the euro area public and private debt securities markets; purchases will be sterilized. | |
| 6/30/2010 | CBPP | Governing Council press release | CBPP finished: Purchases finish on schedule; bonds purchased will be held through maturity. | |
| 10/6/2011 | CBPP2 | Governing Council press release | CBPP2 announced: The ECB will purchase €40 billion in euro-denominated covered bonds. | |
| 12/8/2011 | LTRO | Governing Council press release | LTRO expanded: 36-month LTROs are announced; eligible collateral is expanded. | ECB lowers the main refinancing rate by 0.25% to 1%, and the rate on the marginal lending facility by 0.25% to 1.75%. |
| 8/2/2012 | OMT | ECB press conference | ECB President Mario Draghi indicates that the ECB will expand sovereign debt purchases. He proclaims that "the euro is irreversible." | |
| 9/6/2012 | OMT | Governing Council press release | OMTs announced: Countries that apply to the European Stabilization Mechanism (ESM) for aid and abide by the ESM's terms and conditions will be eligible to have their debt purchased in unlimited amounts on the secondary market by the ECB. | |

***Tab. 8.*** *European Central Bank Important QE Announcements (after Fawley, Neely (2013)).*

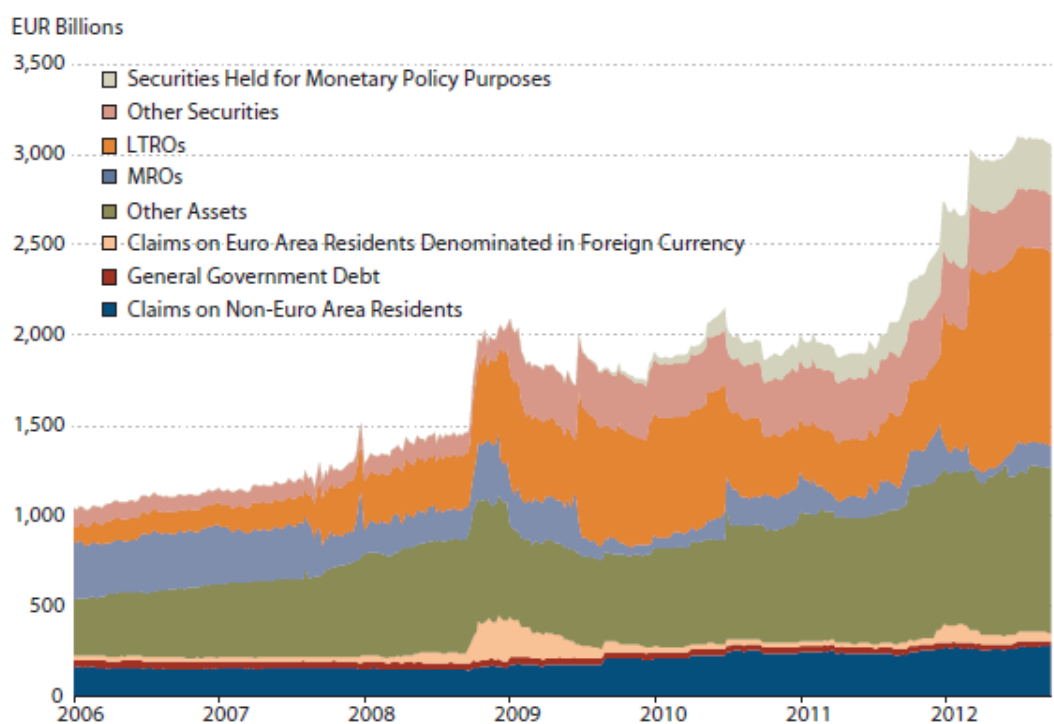

***Fig. 11.*** *European Central Bank assets (after Fawley, Neely (2013)).*

*Fawley, Neely (2013)* present a comparative analysis of the monetary base expansion data and *M2* data in the *USA, European Union, Japan and UK* in Fig. 12.

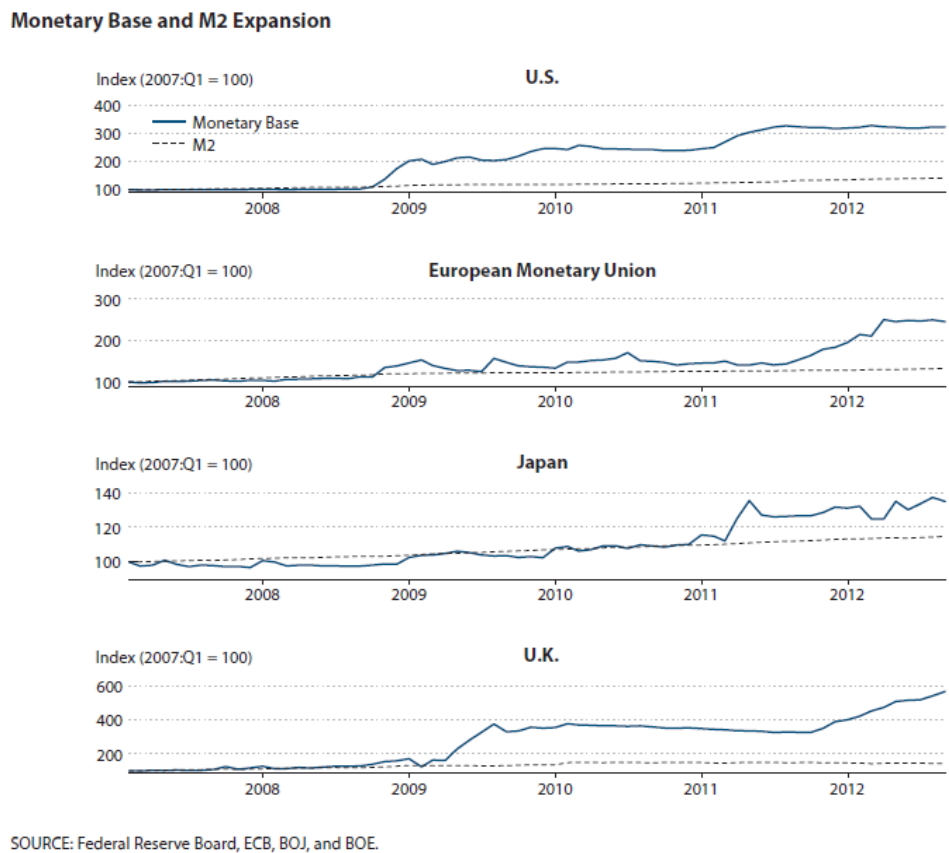

***Fig. 12.*** *Monetary Base and M2 Expansion in USA European Union, Japan and UK (after Fawley, Neely (2013)).*

Finally, let us explain that the various problems on the *QE* policy implementation were comprehensively researched by many other scientists as reviewed in *Fawley, Neely (2013)*: "Academics have already conducted substantial research on recent *QE* programs. *Stroebel and Taylor (2009), Kohn (2009), Meyer and Bomfim (2010),* and *Gagnon et al. (2011a,b),* for example, study the *Fed's 2008-09 QE programs*. *Gagnon et al.'s (2011a,b)* announcement study finds that large-scale asset purchase (*LSAP*) announcements reduced *U.S.* long-term yields. *Joyce et al. (2011)* find that the *BOE's QE program* had bond yield effects quantitatively similar to those reported by *Gagnon et al. (2011a,b)* for the *U.S.* program. *Hamilton and Wu (2011)* indirectly calculate the effects of the *Fed's 2008-09 QE* programs with a term structure model. *Neely (2012)* evaluates the effect of the *Fed's 2008-09 QE* on international long bond yields and exchange rates, showing that the effects are consistent with a simple portfolio balance model and long-run purchasing power parity."

**Discussion on some aspects of quantitative easing policy implementation**

*Sibert (2010)* made the analytic research on the *quantitative easing* policy, showing the *US Federal Reserve System* simplified balance sheet in Tab. 9.

| Assets | Liabilities |
|---|---|
| Securities | Currency in circulation |
| Repos | Depository institutions' balances |
| Loans | Reverse Repos |
| Other Assets | Other Assets |
| | Net worth |

***Tab. 9.*** *US Federal Reserve System simplified balance sheet (after Sibert (2010)).*

*Sibert (2010)* explains the essence of the monetary policy by the *US Federal Reserve System* in the normal times: "In normal times it is typical for modern central banks to make monetary policy by choosing a target short-term policy interest rate. The *Federal Open Market Committee (FOMC)* of the *Federal Reserve System* targets the federal funds rate, the rate at which private deposit-taking institutions lend balances at the *Federal Reserve* overnight to each other. The *Federal Reserve Bank of New York*, acting as agent for the *FOMC*, conducts open-market operations to attain this targeted rate. In usual times this entails offsetting transitory changes in depository institutions' reserves. If the central bank wants to increase these reserves it engages in repurchase agreements ("repos"). These operations are equivalent to short-term collateralized loans but technically they are arrangements where the *Federal Reserve* buys securities in exchange for reserves and agrees to subsequently resell them. As a result, the *Federal Reserve's* balance sheet temporarily expands: the monetary base component of its liabilities rises, as does the repo component of its assets. If the *Federal Reserve* wants to decrease depository institutions' reserves it engages in reverse repos. These are similar to short-term collateralized borrowing but technically they are an arrangement where the *Federal Reserve* sells assets in exchange for reserves and agrees to of the *Federal Reserve's* liabilities: the monetary base component falls and the reverse repo component rises. To offset more permanent factors that would keep the federal funds rate from its target, the *Federal Reserve* must make a more permanent change in its balance sheet. To permanently increase liquidity the *Federal Reserve* expands its balance sheet: purchasing securities and increasing the monetary base. To decrease liquidity it contracts its balance sheet: selling securities and decreasing the monetary base."

*Sibert (2010)* the essence of the monetary policy by the *US Federal Reserve System* in the times of crisis: "Unfortunately for the *FOMC*, as well as other monetary policy committees around the world, by the time the global financial crisis started in earnest in *September 2008*, policy interest rates were already quite low and there was little scope to reduce them further... As seen in Figure 13 below, the federal funds target rate was *2.0* percent in *September 2008*; by *December* of that year it had been reduced to *0 - .25 percent.*

Unable to make monetary policy by announcing lower target interest rates and achieving them through open-market operations, central banks have sought alternatives. ***A potential idea was that instead of announcing a policy interest rate, the central bank would simply engage in further expansionary open-market operations, selling home-currency denominated securities to increase the monetary base.*** This idea, referred to then as ***quantitative easing***, was tried in *Japan* in *2000*."

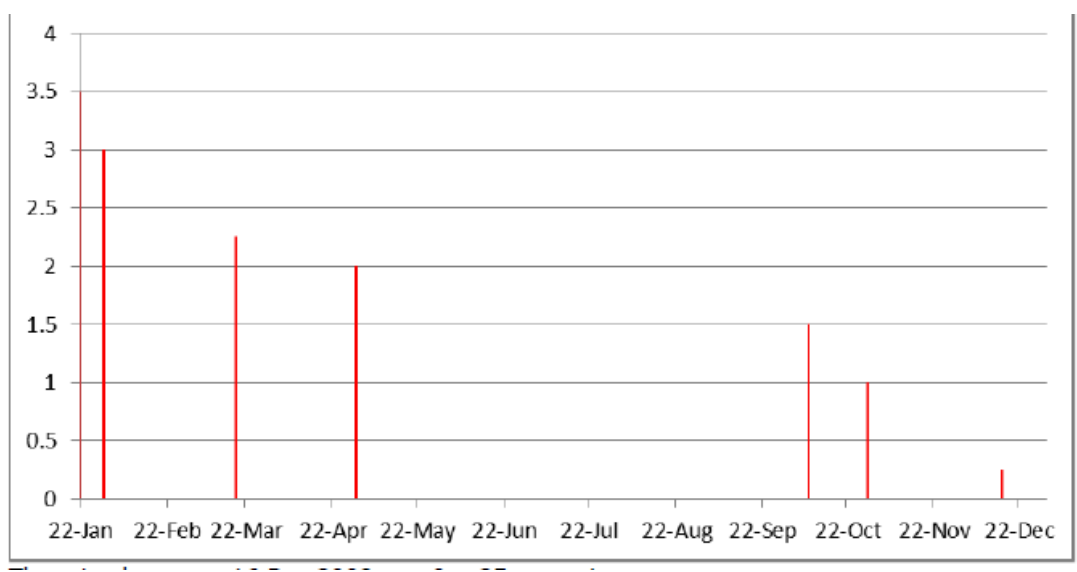

*Fig. 13. US Federal Reserve System target interest rates(after Sibert (2010)).*

*Sibert (2010)* further explains: "In an attempt to free up illiquid markets and deal with failed financial institutions, central banks began to explore more unusual types of monetary policy. In the *United States*, the *Federal Reserve* bought the debt of *Fannie Mae*, *Freddie Mac* and the *Federal Home Loan* banks, as well as mortgage-backed obligations guaranteed by *Fannie Mae*, *Freddie Mac* and *Ginnie Mae*. It also engaged in other crisis-related activities such as the creation of the *Maiden Lanes I, II* and *III* vehicles. As shown in Figure 14 below, between the end of *2007* and the end of *2008* the assets of the *Federal Reserve System* mushroomed from about *$915* billion to *$2,316* billion."

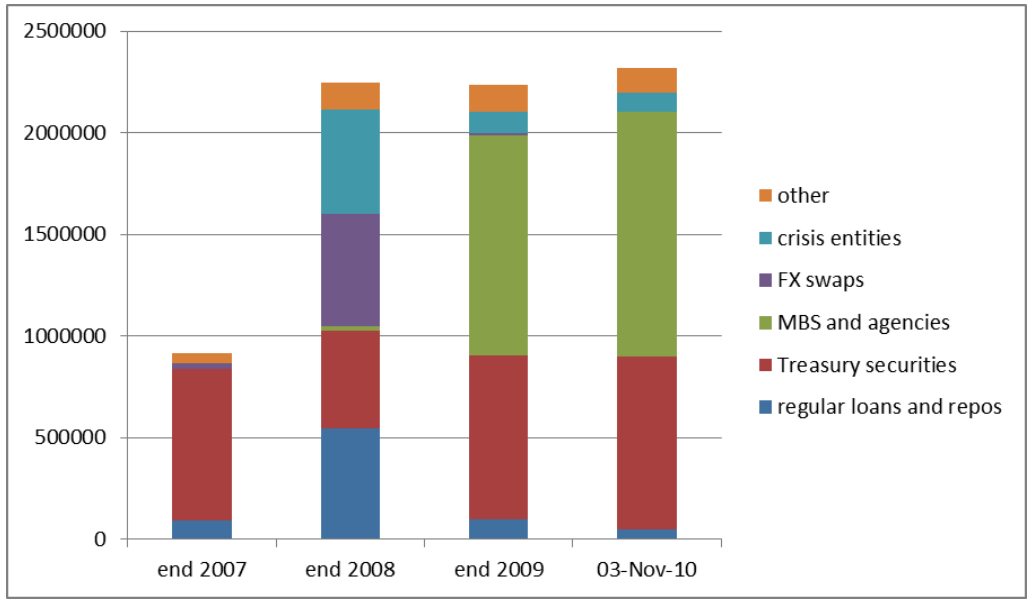

*Fig. 14. US Federal Reserve System assets composition (after Sibert (2010)).*

*Sibert (2010)* attracts some attention to the fact that there are the considerable growing global imbalances between the national economies of various countries as shown in Fig. 15. *Sibert (2010)* writes: "The consternation undoubtedly reflects some genuine concern that *QEII* might be effective and that a lower real value of the dollar will threaten the ability of other countries to pursue export-led growth. However, this is probably not the entire reason: the anger, as well as recent the furor over currency wars is a symptom of the dissatisfaction with global imbalances."

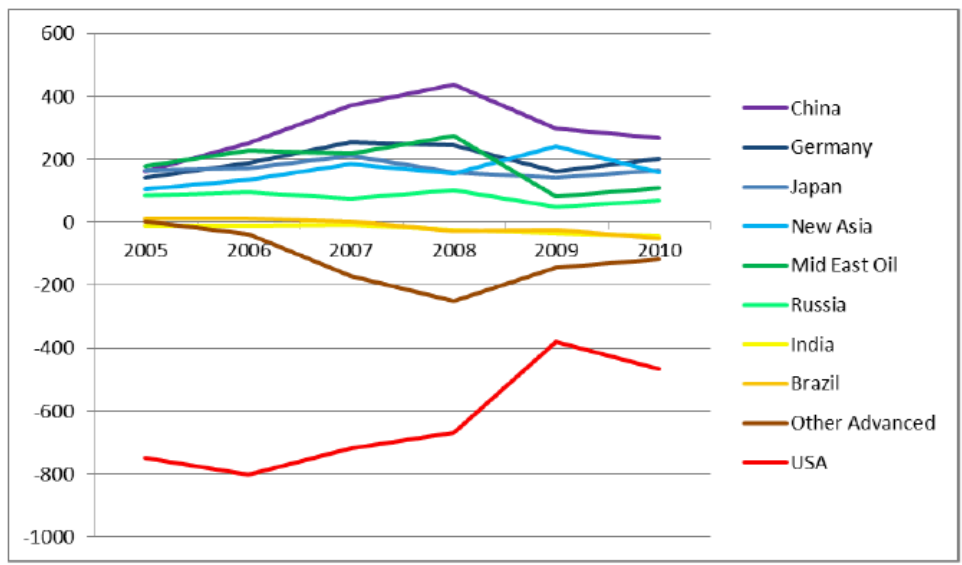

*Fig. 15. Global imbalances, in billions of US dollars (after Sibert (2010)).*

Discussing the *QE* program opportunities, *Bank of England (2009)* explains: "Normally, central banks do not intervene in private sector asset markets by buying or selling private sector debt. But in exceptional circumstances, such intervention may be warranted – for example, when corporate credit markets became blocked as the financial crisis intensified towards the end of *2008*. *Bank of England* purchases of private sector debt can help to unblock corporate credit markets, by reassuring market participants that there is a ready buyer should they wish to sell. That should help bring down the cost of borrowing, making it easier and cheaper for companies to raise finance which they can then invest in their business."

In Fig. 16 the *Bank of England* assets purchases scheme is shown in *Bank of England (2009)*.

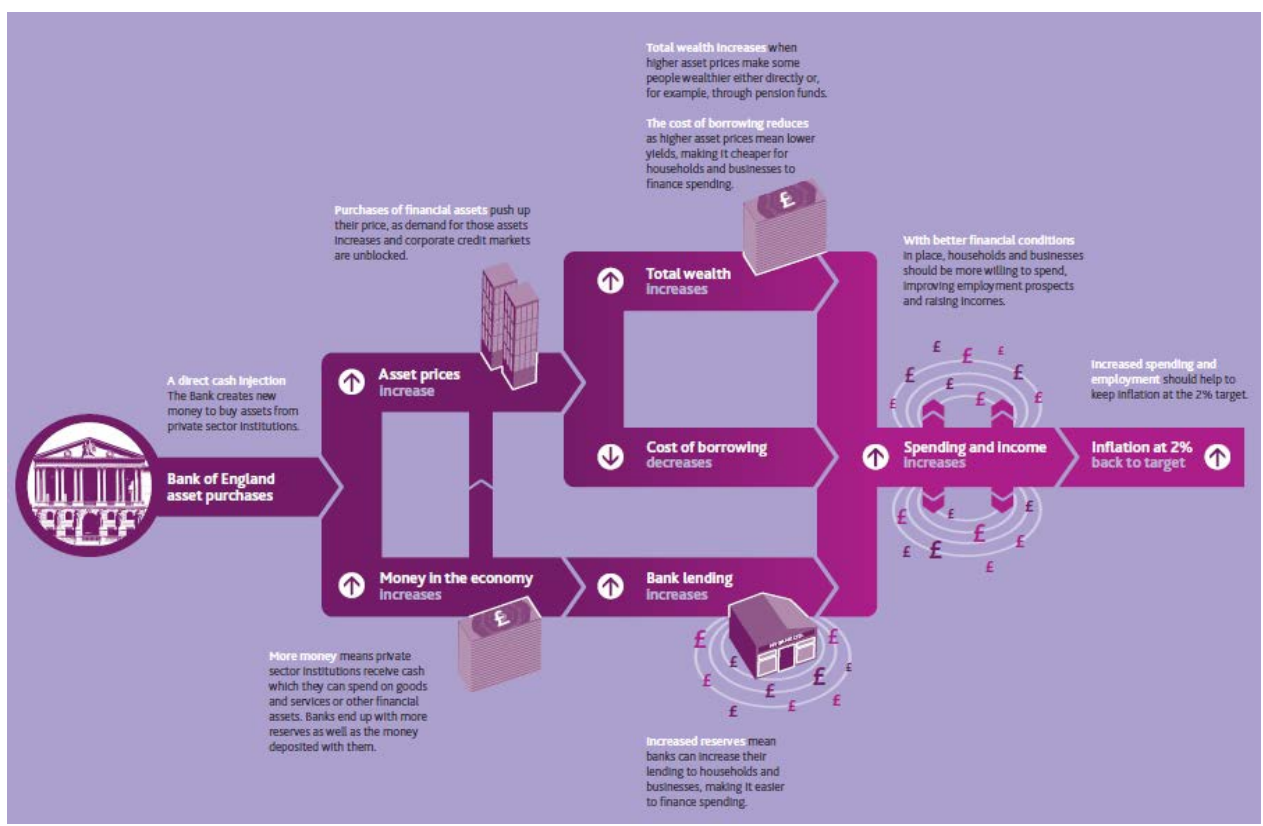

***Fig. 16.*** *Bank of England assets purchases scheme (after Bank of England (2009)).*

*Benford, Berry, Nikolov, Young (2009)* write*:* "The *Bank of England* is the sole supplier of central bank money in sterling. As well as banknotes, central bank money takes the form of reserve balances held by banks at the *Bank of England*. These balances are used to make payments between different banks. The *Bank* can create new money electronically by increasing the balance on a reserve account. So when the *Bank* purchases an asset from a bank, for example, it simply credits that bank's reserve account with the additional funds. This generates an expansion in the supply of central bank money."

*Benford, Berry, Nikolov, Young (2009)* note*:* "Asset purchases are a natural extension of the *Bank's* conventional monetary policy operations. In normal circumstances, the *Bank of England* provides reserves according to the demand from banks at the prevailing level of *Bank Rate*. When conducting asset purchases, the *Bank* is seeking to influence the quantity of money in the economy by injecting additional reserves. This does not mean though that the *Bank* no longer has influence over market interest rates. Market interest rates will continue to be affected both by the level of *Bank Rate* and, in addition, by the amount of reserves that the *Bank* is injecting as investors react to the additional money that they hold in their portfolios."

*Benford, Berry, Nikolov, Young (2009)* explain*:* "Money is highly liquid because it can easily be used to buy goods and services or other assets. The increase in private sector liquidity will depend on the liquidity of the assets that are being exchanged for money. There are a number of channels through which greater liquidity can have an impact. Three key channels are set out below. The transmission mechanism is also summarized in Figure 17."

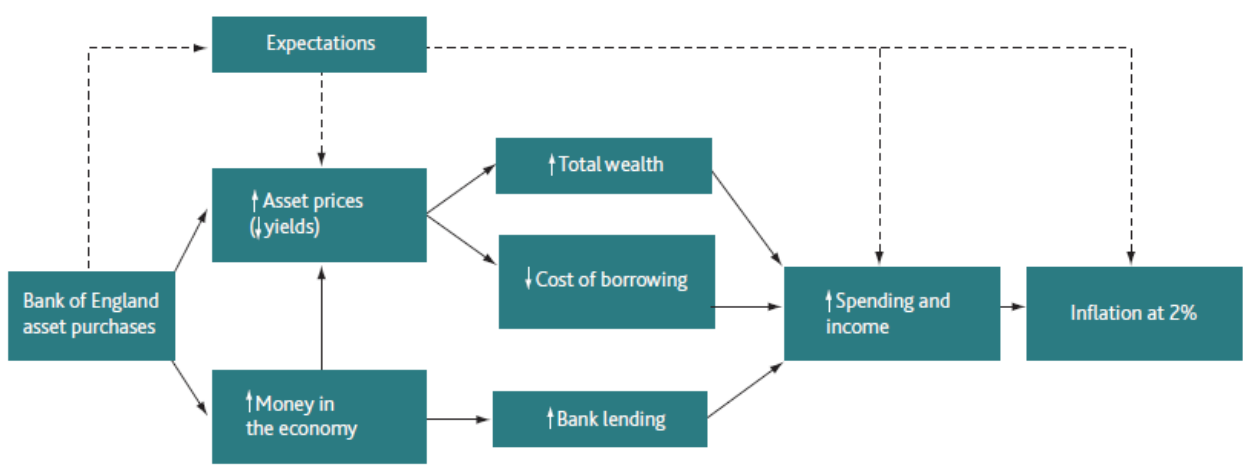

*Fig. 17. Stylized transmission mechanism for asset purchases (after Benford, Berry, Nikolov, Young (2009)).*

*Benford, Berry, Nikolov, Young (2009)* discuss the exit strategy from the *QE* policy: "Monetary policy could be tightened in a number of ways. It could involve some combination of increases in *Bank Rate* and sales of assets in order to reduce the supply of money in the economy. Alternatively, the supply of reserves could be reduced without asset sales, through the issuance of short-term *Bank of England* bills."

*Joyce, Tong, Woods (2011)* made the research on the design, operation and impact of the Bank's asset purchase program that began in *2009* in response to the intensification of the financial crisis. *Joyce, Tong, Woods (2011)* write: "When financial markets are dysfunctional, central bank asset purchases can improve market functioning by increasing liquidity through actively encouraging trading. Asset prices may therefore increase through lower premia for illiquidity. The effects of this channel may, however, only persist while the monetary authority is conducting asset purchases."

In Fig. 18, the **Quantitative Easing (QE) transmission channels** are shown in *Joyce, Tong, Woods (2011)*.

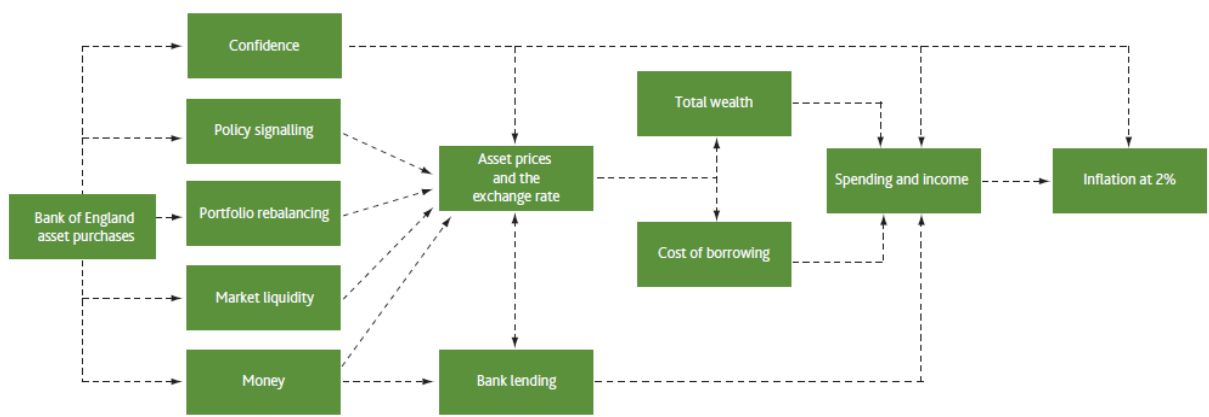

*Fig. 18.* Quantitative Easing (QE) transmission channels (after Joyce, Tong, Woods (2011)).

*Joyce, Tong, Woods (2011)* mention that the estimates of the macroeconomic impact of *QE* can be done, using the following models:

1. **SVAR approach:** A common approach is to characterize *QE* solely through its effects on longer-maturity government bond yields. The simplest starting point for this kind of analysis is to use a small structural vector auto-regression (*SVAR*) containing the policy rate, a government bond yield (the ten-year spot rate), real *GDP* growth and *CPI* inflation. A '*QE* shock' can be identified by assuming that a negative shock to bond yields leads to a contemporaneous rise in *GDP* and *CPI* inflation, but has no effect on policy rates (which are constrained at the zero bound). Estimating this model using quarterly *UK* data over a sample period predating the crisis (*1992 Q1* to *2007 Q2*), and shocking the ten-year gilt yield by *100* basis points, results in a peak impact on the level of real *GDP* of just under *1.5 %* and a peak effect on annual *CPI* inflation of about ¾ percentage points. These effects should be taken as illustrative, given the simplicity of the model and the fact that it has been estimated on a sample predating the crisis. Furthermore, in looking at an impulse response, the assumption is that *QE* is similar to a typical one-period shock to bond yields. This implies rather less persistence than might be expected if *QE* has operated primarily through a portfolio balance effect. Despite these caveats, the effects on key macro variables appear economically significant.

2. *Multiple time-series models approach:* It is also possible to use more sophisticated econometric models to look at this issue. *Kapetanios et al (2011)* use three different time-series models of varying complexity to conduct counterfactual forecasts of the effects of *QE*. The approach (broadly similar in spirit to *Lenza, Pill and Reichlin (2010)*) is to use these models to conduct conditional forecasts under 'policy' and 'no policy' scenarios and then to attribute the difference in the resulting forecasts to the effects of the policy. In the no policy scenario, it is assumed that without *QE* five and ten-year gilt yields would have been *100* basis points higher, although a variety of alternative scenarios are also examined. Averaging across the models suggests that *QE* had a peak effect on the level of *GDP* of around *1 ½ %* and a peak effect on annual *CPI* inflation of about *1 ¼ %* percentage points. These estimates vary considerably across the individual model specifications, and with the assumptions made to generate the counterfactual forecasts, suggesting they are subject to considerable uncertainty.

3. *Monetary approach*: An alternative method of estimating the effects of *QE* is to focus on its impact on the money supply. *Bridges and Thomas (2011)* first calculate the impact of *QE* on the money supply, allowing for the various other influences on broad money over the period. They then apply their estimates to two econometric models — an aggregate *SVAR* model and a linked set of sectoral money demand systems — that allow them to calculate how asset prices and spending need to adjust to make money demand consistent with the increase in broad money supply. Their preferred model estimates suggest that the higher money supply resulting from *QE* may have boosted the level of *GDP* by around *2 %* and *CPI* inflation by about *1%*, though again these estimates are subject to a lot of uncertainty.

4. *Bottom-up approach*: Ideally one would want to make an assessment using a properly specified structural model. But no such model embodying all the relevant transmission channels discussed earlier appears to exist. The forecasting model used by the Bank of England, in common with most large-scale macroeconomic models, does not explain risk premia and therefore does not embody a portfolio balance channel. But, to make a rough calculation, one can take a more bottom-up approach. More specifically, the effects of the QE policy can be broken down into two main elements: (*1*) the impact of asset purchases on gilt prices and other asset prices and (*2*) the effect of asset prices on demand and hence inflation.

A number of ways of estimating (*1*) were already discussed above. The analysis of the *QE* announcement effects suggested that asset purchases pushed down medium to long-term gilt yields by about *100* basis points. The effect of *QE* on a broader range of asset prices is much more uncertain, but there was an immediate *70* basis points fall in investment-grade corporate bond yields and a *150* basis points fall in sub-investment grade yields. There is considerable uncertainty about the effect on equity prices and the immediate market reaction is unlikely to provide an accurate guide, but an estimated portfolio balance model would suggest an impact of around *20 %*. Combining these effects on government and corporate bonds and equity prices suggests an overall boost to households' net financial wealth of about *16 %*.

To quantify the next leg in the transmission mechanism, between asset prices and demand, a range of simple models may be used. To calculate the impact on consumer spending, it is necessary to calculate the wealth elasticity of consumption. One way of doing this is to make an annuity calculation, assuming that households perceive the policy's effects as long-lasting and want to spend their extra wealth evenly over their lifetimes. To calculate the effects on business and dwellings investment, one can use *Q* models, where the incentive to invest depends on the market value of capital relative to its replacement cost. Higher asset prices should raise the market value of capital and reduce the cost of finance, boosting investment spending. Allowing for reasonable uncertainty about the initial impact on asset prices, the result of these sorts of calculations would suggest a peak impact on the level of real *GDP* of between ¾ % and 2 ½ %.

*Tab. 10* shows the estimates of the macroeconomic impact of *QE*, peak impact on the level of output and inflation in *Joyce, Tong, Woods (2011)*.

| Method | Level of GDP (per cent) | CPI inflation (percentage points) |
|---|---|---|
| SVAR | 1½ | ¾ |
| Multiple time-series models average impact[a] | 1½ | 1¼ |
| Monetary approach[b] | 2 | 1 |
| Bottom-up approach | 1½–2½ | ¾–2½ |
| Range across methods[c] | 1½–2 | ¾–1½ |

***Tab. 10.*** *Estimates of the macroeconomic impact of QE, peak impact on the level of output and inflation (after Joyce, Tong, Woods (2011)).*

*Joyce, Tong, Woods (2011)* conclude: "These estimates are clearly highly uncertain, particularly as none of the methods used to produce them fully capture all the likely transmission channels set out earlier, but they do suggest that the effects of *QE* were economically significant."

*Krogstrup, Reynard, Sutter (2012)* make a research supposition that the expansion in reserves following recent quantitative easing programs of the *Federal Reserve* may have affected the long-term interest rates through liquidity effects. *Krogstrup, Reynard, Sutter (2012)* define of the liquidity effect as: "**The liquidity effect is the impact of an expansion of the central bank's liabilities on bond yields, irrespective of the type of asset the central bank buys.**" The estimates suggest that between *January 2009* and *2011*, *10-year US Treasury* yields fell 46-85 basis points as a result of liquidity effects in *Krogstrup, Reynard, Sutter (2012)*. Finally, *Krogstrup, Reynard, Sutter (2012)* come to the conclusion: "… the large scale asset purchases carried out by the *Federal Reserve* may have had liquidity effects due to increases in reserves as well as portfolio balance effects of the changes in the supply of *Treasury* outstanding to the public. Failing to take into account such liquidity effects could lead to an underestimation of the impact of the large scale asset purchase programs on long-term government bond yields. While correlation is not causality, preliminary evidence suggests that reserves and yields were indeed correlated during the *ZLB* period, and that this correlation points to economically important effects. Thus, liquidity effects may have reduced long-term yields by *46* to *85* basis points due to the increase in reserves of about *2.5* percentage points of *GDP* between *January 2009* and *January 2011*."

In addition, let us explain that *Iino, Iyetomi (2012)* researched the *Japanese* transaction network consisting of about *800* thousand firms (nodes) and four million business relations (links) with focus on its modular structure in Fig. 19. *Iino, Iyetomi (2012)* write: "Networks formed by firms through their mutual transactions are a manifestation of economic activities. This is a new way to study economic phenomena emphasizing importance of interaction between economic agents. A number of researches on complex networks have been carried out from a physical point of view. The endeavors encompass development of statistical mechanics methods for quantifying network structure, construction of theoretical models for network formation and visualization of networks based on a physical model." We believe that the capital flows propagation in the *transaction networks channels*, created by the complex interconnected transaction networks between the firms, banks and the *US Federal Reserve System,* may have an influence on the *QE* policy implementation in the *US* financial system.

Let us summarize the above research findings by saying that, in our opinion, the *Quantitative Easing (QE) transmission channels* together with the *transaction networks channels* play an important role during the *QE* policy implementation in the *US* financial system. Therefore, the nonlinearities, originated in the *Quantitative Easing (QE) transmission channels* and the *transaction networks channels* in the financial system, need to be investigated properly.

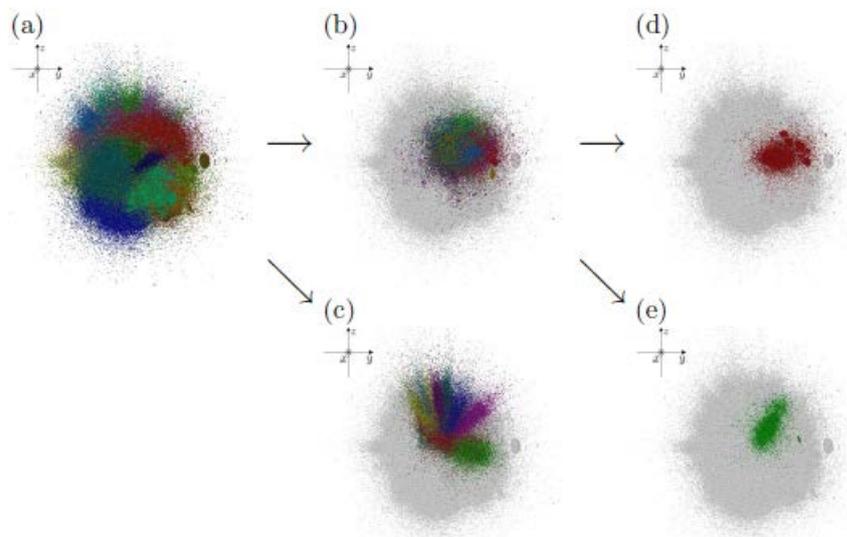

***Fig. 19.*** *The Japanese transaction network drawn in three-dimensional space by a spring-electrical model. Dots in these images represent nodes (firms) whose communities are distinguished by colors. The image (a) shows the whole network. The images (b) and (c) illuminate only the first and second largest communities, respectively. And those communities are further decomposed into sub-communities as displayed with different colors. The first and second largest sub-communities in the largest community are selected in (d) and (e), respectively (after Iino, Iyetomi (2012)).*

**Nonlinear dynamic chaos generation by turbulent capital flows in QE transmission channels and transactions network channels in US financial system**

The modeling of the *USA* economy can be completed with the application of the *VAR* model with the *Smets-Wouters* model, making it possible to provide the reasonably accurate forecasts for the economic series in the selected model *Del Negro, Schorfheide, Smets, Wouters (2007)*, *FRED (2013)*, *Kimball (1995)*, *Lutkepohl (2006)*, *NBER (2013)*, *Smets-Wouters (2002, 2004, 2007)*. The *Smets-Wouters* model is a nonlinear system of equations in the form of a *Dynamic Stochastic General Equilibrium (DSGE)* model that seeks to characterize an economy derived from the economic first principles in *Smets-Wouters (2002, 2004, 2007)*. The considered model works with the *eight* time series: output, prices, wages, hours worked, interest rates, consumption, investment, unemployment in *Matlab (2012)*.

Let us consider the results on the modeling of the *USA* economy, which are shown in Figs. 20 - 26 in *Matlab (2012)*.

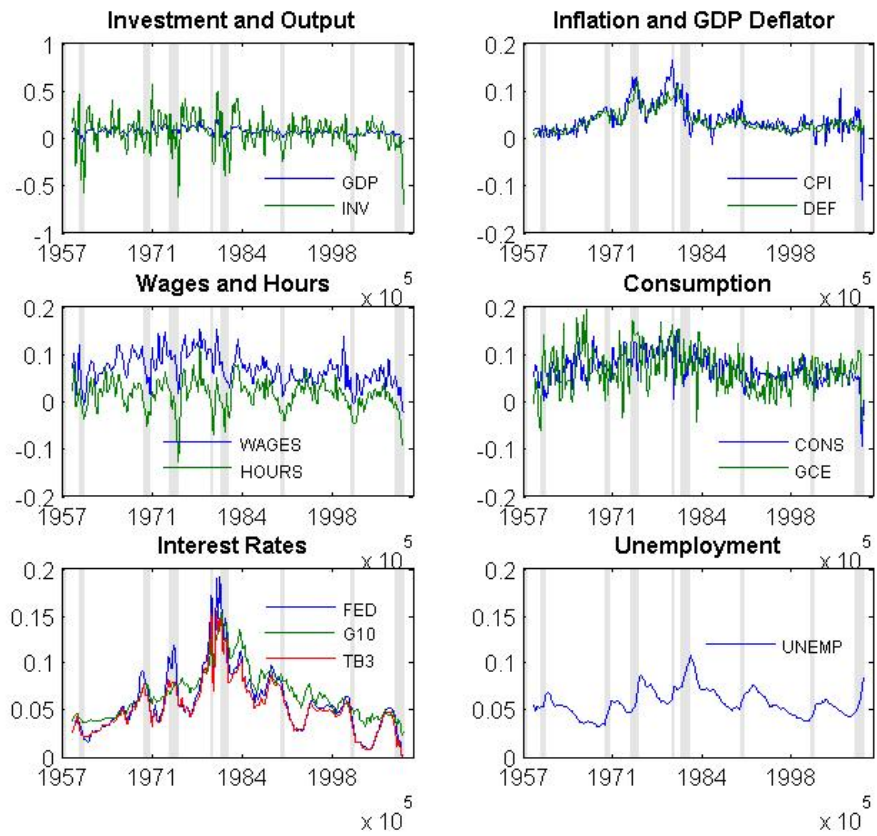

***Fig. 20.*** *Plots of time series with shaded bands that identify periods of economic recession as determined by NBER (after Matlab (2012)).*

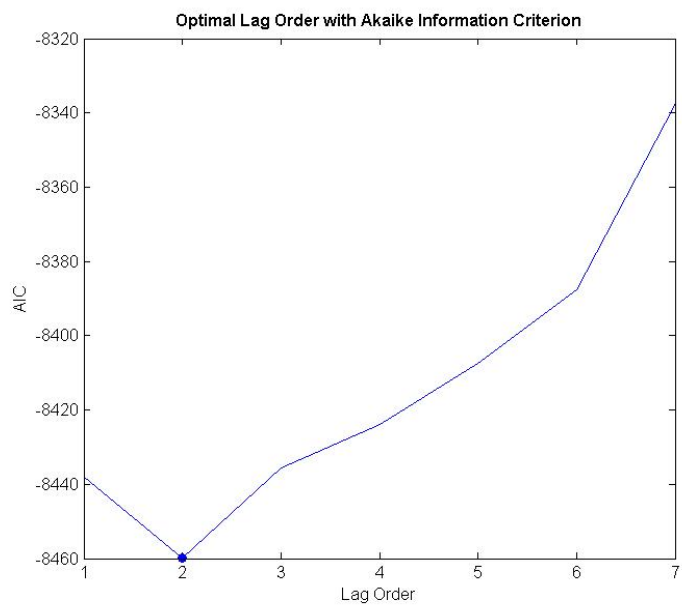

***Fig. 21.*** *"Optimal" number of autoregressive lags based on the Akaike Information Criterion (AIC) (after Matlab (2012)).*

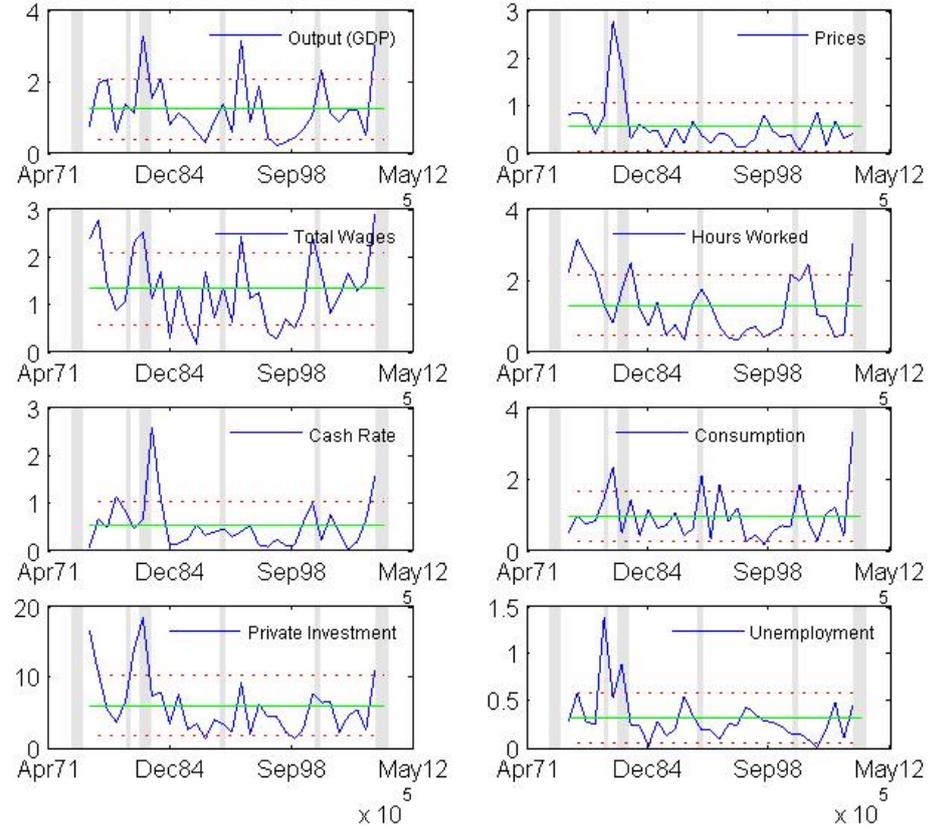

*Fig. 22. Forecast accuracy of model for 1-year horizon (after Matlab (2012)).*

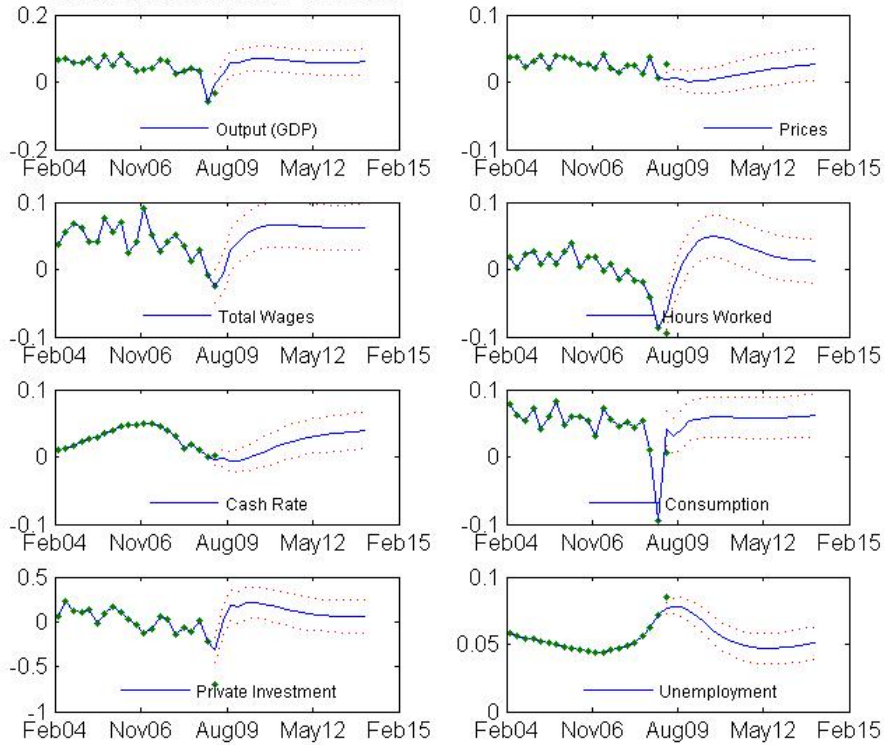

*Fig. 23. Model calibration to December 31, 2008 (after Matlab (2012)).*

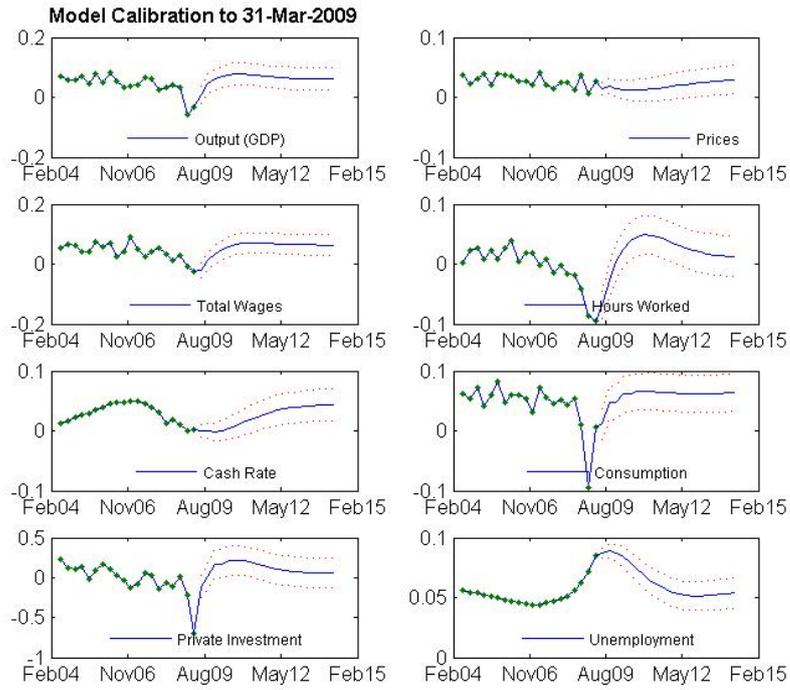

*Fig. 24.* Model calibration to March 31, 2009 (after Matlab (2012)).

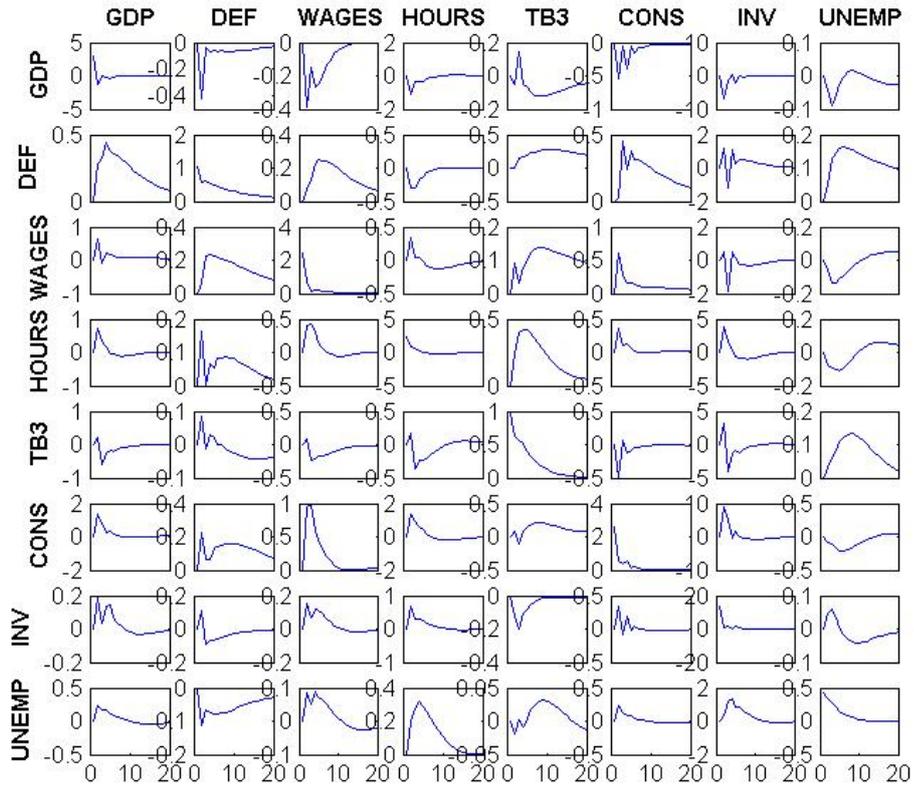

*Fig. 25.* Projected dynamic responses of each time series along each column in reaction to a 1 standard deviation impulse along each row. The units for each plot are percentage deviations from the initial state for each time series (after Matlab (2012)).

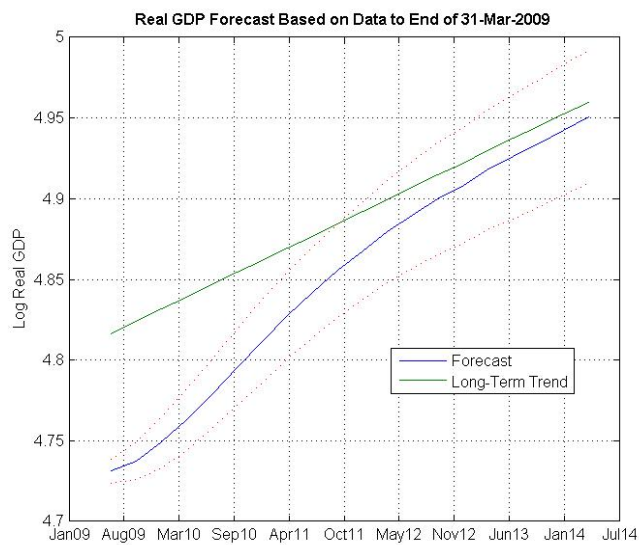

*Fig. 26. Real GDP forecast in the US economy (after Matlab (2012)).*

We would like to comment that the general approach to the modeling of the *US* economy can be based on both:

*1)* the "large" empirically-motivated regression model, and

*2)* the *Value at Risk (VAR)* model with the incorporated *Smets-Wouters* model.

Let us note that the above two approaches don't take to the account an origination of the nonlinear dynamic chaos during the capital flow in the quantitative easing transmission channels and the transaction networks channels in the *US* financial system at the time of the *QE* policy implementation by the *US Federal Reserve System*.

Therefore, we completed the computer modeling, using both the *Nonlinear Dynamic Stochastic General Equilibrium Theory (NDSGET) and the Hydrodynamics Theory (HT)*, to accurately characterize the *US* economy in the conditions of the *QE* policy implementation by the *US Federal Reserve*, because, in our opinion, the **nonlinear dynamic chaos** can be generated by the turbulent capital flows in the *QE* transmission channels in the *US* financial system in *Kolmogorov (1941), Landau (1944, 2001), Feigenbaum (1979, 1980).* In this connection, we focused our attention on a set of the important financial questions in *(see the book on the nonlinearities in microwave superconductivity in D O Ledenyov, V O Ledenyov (2012))*:

1. How does the laminar – turbulent transition in the *QE* transmission channels and the transaction networks channels in the *US* financial system occur?
2. How does the inflow of capital with the turbulent spectrum in the *QE* transmission channels and the transaction networks channels impact the macroeconomics situations in the *US* economy?

3. How does the inflow of capital with the turbulent spectrum influence the business cycle's characteristics during the business cycle's propagation in and interaction with the *US* financial and economic systems?
4. Will the degree of nonlinearity of the *US* financial and economic systems increase as a result of the increase of capital flows with the turbulent spectrum in the *QE* transmission channels during the *QE* policy introduction and implementation by the *US Federal Reserve System*?

Aiming to clarify the above research problems, we developed the two complex recursive algorithms and applied the parallel computing techniques to complete the computer modeling with the goal to accurately characterize the impulse responses by the *US* financial and economic systems to the changing levels of liquidity due to the *QE* program implementation. For example, we completed the following computer simulations:

*1)* the dependence of the capital flows propagation in the quantitative easing transmission channels in the *US* financial system over the specified time period, and

*2)* the dependence of the capital flows propagation in the transaction networks channels, created by the complex interconnected transaction networks between the firms, banks and the *US Federal Reserve System, in the US financial system* over the certain time period.

In our research, the cross-sections of both the quantitative easing transmission channels and the transaction networks channels have the complex dependencies on a number of the financial and economic variables in the *US* financial system. Using the knowledge base in the econophysics, we assumed that the *nonlinear dynamic chaos* can be generated by the turbulent capital flows in both the quantitative easing transmission channels and the transaction networks channels in the *US* financial system, when there are the laminar - turbulent capital flows transitions. We demonstrated that the capital flows in the quantitative easing transmission channels and the transaction networks channels in the *US* financial system can be accurately characterized by the *Reynolds* numbers in *Reynolds (1883)*:

$$\mathrm{Re} = \frac{vd}{\upsilon},$$

where $d$ is the characteristic geometric dimension of researched system, $v$ is the coefficient of kinematic viscosity, $\upsilon$ is the characteristic velocity of capital flow. We think that the transition to the nonlinear dynamic chaos regime of the *US* financial system operation can be realized through the cascade of the *Landau – Hopf* bifurcations in the turbulent capital flows in both the quantitative easing transmission channels and the transaction networks channels in the *US* financial system in *Landau (1944, 2001), Hopf (1948)*. We found that the ability of the *US* financial system to adjust to the different levels of liquidity strongly depends on the nonlinearities appearance in the *QE* transmission channels, and it is limited by the laminar – turbulent capital flows transitions in the *QE* transmission channels and the transaction networks channels in the *US* financial system. In our opinion, one of the interesting modeling results is that the main purpose of liquidity adding into the US financial system is to stabilize the *US* financial system; however, we found that the instabilities may appear at the certain levels of added liquidity in the *QE* transmission channels in the *US* financial system, because of the laminar – turbulent capital flows transitions, resulting in the switching to the nonlinear dynamic chaos regime of the *US* financial system operation. Therefore, we made an important conclusion that the *QE* program implementation may stabilize or destabilize the *US* financial system, depending on the level of added liquidity in the *US* financial system over some time period.

## Conclusion

The central banks introduced a series of quantitative easing programs and decreased the long term interest rates to near zero with the aim to ease the credit conditions and provide the liquidity into the financial systems, responding to the *2007-2013* financial crisis in the *USA*, *UK*, *Western Europe, and Japan*. We reviewed the *U.S. Federal Reserve System, European Central bank, Bank of England and Bank of Japan* monetary and financial policies with the particular focus on the quantitative easing policy implementation in the *USA*. Discussing some aspects of the quantitative easing policy implementation, we highlighted the fact that the levels of capital may change quickly in the quantitative easing transmission channels and in the transaction networks channels during the quantitative easing policy implementation, when the liquidity is added to the financial system. In agreement with the recent research findings in the econophysics, we propose that the nonlinear dynamic chaos can be generated by the turbulent capital flows in quantitative easing transmission channels and in the transaction networks channels, when the laminar capital flows transform to the turbulent capital flows in the *US* financial system. We demonstrated that the capital flows in both the quantitative easing transmission channels and the transaction networks channels in the *US* financial system can be accurately characterized by the *Reynolds* numbers. We explained that the transition to the nonlinear dynamic chaos regime can be realized through the cascade of the *Landau – Hopf* bifurcations during the turbulent capital flows in both the quantitative easing transmission channels and the transaction networks channels in the *US* financial system. Finally, we clarify that the general approach to the modeling of the *US* economy is based on both *1)* the "large" empirically-motivated regression model, and *2)* the *Value at Risk (VAR)* model with the incorporated *Smets-Wouters* model, which don't take to the account the origination of the nonlinear dynamic chaos regime during the capital flow in the quantitative easing transmission channels and in the transaction networks channels in the *US* financial system. Therefore, we completed the computer modeling, using both the *Nonlinear Dynamic Stochastic General Equilibrium Theory (NDSGET) and the Hydrodynamics Theory (HT)*, to accurately characterize the *US* economy in the conditions of the *QE* policy implementation by the *US Federal Reserve System*. We found that the ability of the *US* financial system to adjust to the different levels of liquidity strongly depends on the nonlinearities appearance in the *QE* transmission channels, and is limited by the laminar – turbulent capital flow transitions in the *QE* transmission channels at the *US* financial system operation. The developed computer model allows us to make the very accurate long- and short- time forecasts of the *US* economy performance in the cases, when the different levels of liquidity are added into the *US* financial system.

## Acknowledgement

This research article outlines the existing problems and provides some insightful ideas on the possible solutions to the *US* financial system problems, appearing during the quantitative easing policy implementation by the US Federal Reserve System. Authors would like to thank Dr. Ben Shalom Bernanke, *Chairman of the Board of Governors of the Federal Reserve System* for his brilliant thoughts, interesting scientific discussions and valuable practical advices on the quantitative easing policy implementation by the *US Federal Reserve System* in the U.S.A. It makes sense to highlight the fact that our research proposal that the financial system is dynamic, hence the financial variables, characterizing the stability of the *US* financial system, have to be taken to the account during the *US* financial system stability monitoring and its vulnerabilities evaluation (see the research article: "On the accurate characterization of business cycles in nonlinear dynamic financial and economic systems" by Dimitri O. Ledenyov and Viktor O. Ledenyov 1304.4807.pdf www.arxiv.org on April 13, 2013) was fully supported in the speech: "Monitoring the Financial System" by Dr. Ben Shalom Bernanke at the *49th Annual Conference on Bank Structure and Competition* in Chicago, Illinois in the U.S.A. on May 10, 2013. In addition, we thank the *Federal Reserve System* for giving us an opportunity to analyse the requested research articles, reports, reviews and financial data on the *QE* policy implementation in the *USA*. Authors have a strategic vision that more advanced research on the quantitative easing policy implementation is required to improve our understanding of the problem, taking to the consideration the complexity of conducted computer modeling and the novelty of obtained research results.

*E-mail: dimitri.ledenyov@my.jcu.edu.au .